\documentclass[twocolumn]{bmcart}% uncomment this for twocolumn layout and comment line below
\usepackage[utf8]{inputenc} %unicode support

\usepackage{amsmath}
\usepackage{amssymb}
\usepackage{booktabs}
\usepackage{graphicx}
\usepackage{caption}
\usepackage{subcaption}
\usepackage{algorithm}
\usepackage{algpseudocode}
\usepackage{rotating}
\usepackage{hyperref}

\startlocaldefs
\DeclareMathOperator*{\argmax}{arg\,max}
\DeclareMathOperator*{\argmin}{arg\,min}
\numberwithin{equation}{section}
\def\beginnew{}
\def\endnew{}
\endlocaldefs

\begin{document}

%%% Start of article front matter
\begin{frontmatter}

\begin{fmbox}
\dochead{Research}


%\title{A sample article title}
\title{A Machine-Learning Phase Classification Scheme for Anomaly Detection in Signals with Periodic Characteristics}

%%%%%%%%%%%%%%%%%%%%%%%%%%%%%%%%%%%%%%%%%%%%%%
%%                                          %%
%% Enter the authors here                   %%
%%                                          %%
%% Specify information, if available,       %%
%% in the form:                             %%
%%   <key>={<id1>,<id2>}                    %%
%%   <key>=                                 %%
%% Comment or delete the keys which are     %%
%% not used. Repeat \author command as much %%
%% as required.                             %%
%%                                          %%
%%%%%%%%%%%%%%%%%%%%%%%%%%%%%%%%%%%%%%%%%%%%%%

%\author[
%   addressref={aff1},                   % id's of addresses, e.g. {aff1,aff2}
%   corref={aff1},                       % id of corresponding address, if any
%   noteref={n1},                        % id's of article notes, if any
%   email={jane.e.doe@cambridge.co.uk}   % email address
%]{\inits{JE}\fnm{Jane E} \snm{Doe}}
%\author[
%   addressref={aff1,aff2},
%   email={john.RS.Smith@cambridge.co.uk}
%]{\inits{JRS}\fnm{John RS} \snm{Smith}}

\author[
   addressref={aff1},                   % id's of addresses, e.g. {aff1,aff2}
   %corref={aff1},                      % id of corresponding address, if any
   %noteref={n1},                       % id's of article notes, if any
   email={Lia.Ahrens@dfki.de}           % email address
]{\inits{L}\fnm{Lia} \snm{Ahrens}}
\author[
   addressref={aff1},                   % id's of addresses, e.g. {aff1,aff2}
   %corref={aff1},                      % id of corresponding address, if any
   %noteref={n1},                       % id's of article notes, if any
   email={Julian.Ahrens@dfki.de}        % email address
]{\inits{J}\fnm{Julian} \snm{Ahrens}}
\author[
   addressref={aff1,aff2},              % id's of addresses, e.g. {aff1,aff2}
   %corref={aff1},                      % id of corresponding address, if any
   %noteref={n1},                       % id's of article notes, if any
   email={Hans_Dieter.Schotten@dfki.de} % email address
]{\inits{HD}\fnm{Hans D.} \snm{Schotten}}

%%%%%%%%%%%%%%%%%%%%%%%%%%%%%%%%%%%%%%%%%%%%%%
%%                                          %%
%% Enter the authors' addresses here        %%
%%                                          %%
%% Repeat \address commands as much as      %%
%% required.                                %%
%%                                          %%
%%%%%%%%%%%%%%%%%%%%%%%%%%%%%%%%%%%%%%%%%%%%%%

%\address[id=aff1]{%                           % unique id
%  \orgname{Department of Zoology, Cambridge}, % university, etc
%  \street{Waterloo Road},                     %
%  %\postcode{}                                % post or zip code
%  \city{London},                              % city
%  \cny{UK}                                    % country
%}
%\address[id=aff2]{%
%  \orgname{Marine Ecology Department, Institute of Marine Sciences Kiel},
%  \street{D\"{u}sternbrooker Weg 20},
%  \postcode{24105}
%  \city{Kiel},
%  \cny{Germany}
%}

\address[id=aff1]{%                           % unique id
  \orgname{Deutsches Forschungszentrum f\"{u}r K\"{u}nstliche Intelligenz}, % university, etc
  \street{Trippstadter Stra\ss e 112}         %
  \postcode{67663}                            % post or zip code
  \city{Kaiserslautern},                      % city
  \cny{Germany}                               % country
}
\address[id=aff2]{%                           % unique id
  \orgname{Technische Universit\"{a}t Kaiserslautern}, % university, etc
  \street{Paul-Ehrlich-Straße 11}             %
  \postcode{67663}                            % post or zip code
  \city{Kaiserslautern},                      % city
  \cny{Germany}                               % country
}

%%%%%%%%%%%%%%%%%%%%%%%%%%%%%%%%%%%%%%%%%%%%%%
%%                                          %%
%% Enter short notes here                   %%
%%                                          %%
%% Short notes will be after addresses      %%
%% on first page.                           %%
%%                                          %%
%%%%%%%%%%%%%%%%%%%%%%%%%%%%%%%%%%%%%%%%%%%%%%

\begin{artnotes}
%\note{Sample of title note}     % note to the article
%\note[id=n1]{Equal contributor} % note, connected to author
\end{artnotes}

%\end{fmbox}% comment this for two column layout

%%%%%%%%%%%%%%%%%%%%%%%%%%%%%%%%%%%%%%%%%%%%%%
%%                                          %%
%% The Abstract begins here                 %%
%%                                          %%
%% Please refer to the Instructions for     %%
%% authors on http://www.biomedcentral.com  %%
%% and include the section headings         %%
%% accordingly for your article type.       %%
%%                                          %%
%%%%%%%%%%%%%%%%%%%%%%%%%%%%%%%%%%%%%%%%%%%%%%

\begin{abstractbox}

\begin{abstract} % abstract
%\parttitle{First part title} %if any
%Text for this section.
%
%\parttitle{Second part title} %if any
%Text for this section.
In this paper we propose a novel machine-learning method for anomaly detection applicable to data with periodic characteristics where randomly varying period lengths are explicitly allowed.
A multi-dimensional time series analysis is conducted by training a data-adapted classifier consisting of deep convolutional neural networks performing phase classification.
The entire algorithm including data pre-processing, period detection, segmentation, and even dynamic adjustment of the neural networks is implemented for fully automatic execution.
The proposed method is evaluated on three example datasets from the areas of cardiology, intrusion detection, and signal processing, presenting reasonable performance.

\end{abstract}

%%%%%%%%%%%%%%%%%%%%%%%%%%%%%%%%%%%%%%%%%%%%%%
%%                                          %%
%% The keywords begin here                  %%
%%                                          %%
%% Put each keyword in separate \kwd{}.     %%
%%                                          %%
%%%%%%%%%%%%%%%%%%%%%%%%%%%%%%%%%%%%%%%%%%%%%%


% MSC classifications codes, if any
%\begin{keyword}[class=AMS]
%\kwd[Primary ]{}
%\kwd{}
%\kwd[; secondary ]{}
%\end{keyword}

\end{abstractbox}
\end{fmbox}% uncomment this for twcolumn layout

\end{frontmatter}

%%%%%%%%%%%%%%%%%%%%%%%%%%%%%%%%%%%%%%%%%%%%%%
%%                                          %%
%% The Main Body begins here                %%
%%                                          %%
%% Please refer to the instructions for     %%
%% authors on:                              %%
%% http://www.biomedcentral.com/info/authors%%
%% and include the section headings         %%
%% accordingly for your article type.       %%
%%                                          %%
%% See the Results and Discussion section   %%
%% for details on how to create sub-sections%%
%%                                          %%
%% use \cite{...} to cite references        %%
%%  \cite{koon} and                         %%
%%  \cite{oreg,khar,zvai,xjon,schn,pond}    %%
%%  \nocite{smith,marg,hunn,advi,koha,mouse}%%
%%                                          %%
%%%%%%%%%%%%%%%%%%%%%%%%%%%%%%%%%%%%%%%%%%%%%%

%%%%%%%%%%%%%%%%%%%%%%%%% start of article main body
% <put your article body there>

\section{Introduction}

Many real-world systems, both natural and anthropogenic, exhibit periodic behaviour.
Monitoring such systems necessarily produces periodic time series.
In one particular instance of such a monitoring application, one is interested in automatically detecting changes in the periodically repeating pattern and thus anomalies in the systems operation.
This type of anomaly detection occurs in a wide range of different fields and applications, be they medical, e.g.~diagnosing diseases of the cardiovascular and respiratory systems, in industrial contexts, e.g.~monitoring the operation of a transformer or rotating machinery, and in signal processing and communications.
The pursued aims range from simple monitoring to intrusion detection and prevention.

Traditionally, anomaly detection is performed in the form of outlier detection in mathematical statistics.
Numerous methods have been proposed, including but not limited to distance- and density-based techniques~\cite{DNL15,BKNS00} and subspace- or submanifold-based techniques~\cite{KKK04,CSAT17,XCZ18}.
Most of these approaches make no explicit use of the concept of time and are therefore usually less suited for the analysis of time series.
Methods making explicit use of the temporal structure include classical models from statistical time series analysis such as autoregressive--moving average (ARMA) models~\cite{Lou08}, Kalman filters~\cite{TTS07} or more general hidden Markov models~\cite{YSZ03}, and rolling-window distance-based methods such as matrix profiles~\cite{YZU16}.
Distance analysing methods are effective for clean data but not robust against noise, whereas distribution-based methods from mathematical statistics are still powerful in the presence of noise, requiring data-specific parametrisation.
In the past few years, non-linear methods, such as different types of recurrent neural networks (RNNs) and in particular long short-term memory (LSTM) networks, have also come into use~\cite{GSC00,YO17}.
Many of these methods are difficult to train~\cite{BSF94,HBFS01,PMB13} or need large amounts of data in order to achieve reasonable performance while avoiding overfitting.
On the other hand, in recent years, convolutional neural networks (CNNs) have gained popularity in image processing~\cite{KSH12,SIVA16} where they are used mainly for classification tasks.
The same principles that are responsible for the success of CNNs in image processing carry over to other types of signal processing when the number of dimensions of the convolutional kernels is changed accordingly.
Most of the work using recurrent or convolutional networks for time series analysis focusses on forecasting or detecting certain patterns explicitly known at training time.
On these tasks, convolutional networks have recently been shown to outperform the previously state of the art LSTMs~\cite{BBO18}.

In this paper we consider data with periodic characteristics and design a machine-learning algorithm for time series analysis, in particular anomaly detection, applying convolutional neural nets in a manner which, to the best of the authors knowledge, has not been proposed previously.
In contrast to existing methods and inspired by machine-learning methods for image processing, we employ a convolutional net acting not as a predictor or estimator but as a classifier whose classes indicate phase i.e.~the relative location in time.
We also integrate general procedures for data pre-processing and automated phase reclustering so that no manual action is required in between.

Our algorithm is tested on three datasets: a cardiology dataset (ECG database)~\cite{BKS95}, an industrial network dataset for cyber attack research (SCADA dataset)~\cite{LF16}, and a synthetic waveform dataset described in detail in Section~\ref{sec: waves}.
It turns out that, to a certain extent, our method is robust against unclean data and the related neural networks do not show high sensitivity to the hyperparameters and are relatively easy to train.

\beginnew
The remainder of the paper is organised as follows.
In Section~\ref{sec:preliminaries} we specify the types of anomaly detection considered in this paper, comment on traditional methods, and introduce the concept of our solution.
In Section~\ref{sec:concept} our general approach to the considered anomaly detection problems is described in detail, including data pre-processing, mathematical basis of convolutional neural networks, and training algorithm.
In Section~\ref{sec:example datasets} our method is fine-tuned for the three aforementioned example datasets and in Section~\ref{sec:results} the empirical results are evaluated.
In Appendix~\ref{pds}, dealing with the issue of randomly varying period length which shows up in many real world applications such as in the ECG data (Section~\ref{sec:cardio}) and synthetic waves (Section~\ref{sec: waves}), an auxiliary period detection scheme is designed based on classical principles of signal processing.
In Appendix~\ref{sec:comp} we perform some comparisons with other methods for anomaly detection in order to further highlight the advantage of using a convolutional neural network in the proposed manner.
\endnew

\beginnew
\section{Preliminaries}\label{sec:preliminaries}

In preparation for the detailed description of our machine-learning phase classfication scheme given in Section~\ref{sec:concept}, in this section we clarify the tasks of anomaly detection in time series with periodic characteristics, discuss some common methods, and outline the essential ideas of our approach.

\subsection{Context of this work}
In general, a time series $\{X_t\}_{t=0,1,2,\ldots}$ (i.e.~a temporal sequence of observations $X_0,X_1,X_2,\ldots$, also termed signal) is said to exhibit periodic behaviour with \emph{period length} $s$ if similarities occur after every $s$ time units, i.e., observations that are $s$ time units apart, $X_{t_0}, X_{t_0+s},X_{t_0+2s},\ldots$ for any $t_0$, are similar.

Periodic signals occur naturally in a wide range of applications and in a large number of fields such as audio processing, vibration analysis, biomedical engineering, climatology, and economic time series analysis.
Oftentimes, one wishes to monitor the behaviour of such a system.
In particular, a common task when observing a signal is that of \emph{anomaly detection}, i.e., the detection of deviations from a certain normal mode of operation.
This has a variety of applications such as disease diagnosis, network security, fraud recognition in bank transactions, etc.

The general approach to anomaly detection is to relate a mathematical model (parametric or non-parametric) to the normal behaviour of the underlying system based on historic observations (training data) and set a confidence region for data of normal type; applying the data-adapted model to the ongoing observations (test data), whether the output lies within or outside the pre-defined confidence region decides if the corresponding input observation is considered normal or abnormal, respectively.

As with most naturally occurring siganals, many of the aforementioned signals do not satisfy the exact mathematical definition of periodicity.
Instead, they exhibit a property which is referred to as quasiperiodicity which basically means that the signal does not exactly repeat itself, but has deviations both in its values and in the length of the actual periods.
This behaviour is very common for instance in biological or climatological systems.
As a consequence for the task of anomaly detection, a sophisticated mathematical model is required to capture the essence of the diverse and noise corrupted signals.

\subsection{Tasks of anomaly detection}\label{sec:task}
Mathematically, the approach to anomaly detection proposed in this paper applies to the following two types of problems:

\begin{description}
  \item[Type~A]
    The historic observations of normal type (training data) are made up of various signals $\{\{X^{(\iota)}_t\}_t \mid \iota\in I \text{ (index set)} \}$.
    The signals $\{X^{(\iota)}_t\}$, $\iota\in I$, share certain common normal-type-characterising features, but differ in their values and exhibit periodic characteristics with individual period length $s^{(\iota)}$ which may also fluctuate over time.
    The task of anomaly detection in such a setting consists in rating each ongoing observation signal $\{X_t\}_t$ as normal or abnormal.

  \item[Type~B]
    The historic observations of normal type (training data) are made up of consecutive single data points $X_0,X_1,\ldots,X_{N-1}$ which jointly form a time series $\{X_t\}_{t=0,\ldots,N-1}$.
    The occurance of the data points $X_0,\ldots,X_{N-1}$ follows certain normal-type-characterising patterns, which is reflected in the corresponding time series $\{X_t\}_{t=0,\ldots,N-1}$ as seasonal effects associated with period length $s$ where $s$ may randomly vary over time.
    The task of anomaly detection in such a setting consists in specifying segments of onging observations $\{X_{t}\}_{t\geq N}$ which are abnormal.
\end{description}

Problems of Type~A arise from areas such as disease diagnosis, climatology, vibration analysis, etc., whereas problems of Type~B are often addressed in the security sector and building monitoring systems within the framework of signal processing.
In general, establishing an adequate mathematical model for the normal behaviour of a system requires a proper amount of training data.
In our experiments in Section~\ref{sec:example datasets}, our approach to the considered problems is applied to a cardiology dataset for detecting heart disease (cf.~ECG database \cite{BKS95}) as an example of problems of Type~A, a relatively small industry dataset in the context of network security (cf.~SCADA dataset \cite{LF16}) as an example of problems of Type~B, and a more extensive synthetic waveform dataset injected with a variety of noise and anomalies (cf.~Section~\ref{sec: waves}) again as an example of problems of Type~B.
The experimental results are provided in Section~\ref{sec:results}.

From a mathematical perspective, problems of\linebreak Type~A are more challenging than those of Type~B.
In the setting of Type~A a considerably complex mathematical model is needed for capturing diverse variations of the normal behaviour across a variety of training signals, whereas in the setting of Type~B the required mathematical model for the normal behavior is to be fitted to a single training time series.
Many traditional methods for anomaly detection in periodic signals may find direct applications to problems of Type~B but fail to be applicable to problems of Type~A.
This will be further discussed in the subsequent section.

\subsection{On common methods}
Let us comment on the adequacy of some traditional methods for detecting anomalies in periodic signals in our context.

\subsubsection{Distance-analysing methods}
The most straightforward treatment of seasonal data goes back to cross correlation analysis, e.g.~matrix profiles~\cite{YZU16}.
The basic idea therein is to apply a rolling window and define a Euclidean-type metric which measures the distance of consecutive values within the rolling window at different locations of the underlying time series from one another or from a fixed reference sequence (e.g.~a mean window consisting of seasonal means); data points exhibiting large distance from the reference value are considered abnormal.

In general, distance-analysing approaches are not resistant against noise and fail to capture complex structures in the data.
In the appendix, we evaluate a simple distance-based self-similarity approach in Section~\ref{sec:comp-self-sim}.
We also provide a distance-based version of our phase classification scheme (without artificial neural networks) for comparison in Section~\ref{sec:comp-dist-cls}.

\subsubsection{ARIMA methodology and Kalman filtering}\label{sec:ARIMA}
A more sophisticated class of methods arises from mathematical statistics, e.g.~autoregressive integrated moving average (ARIMA) methodology, methods\linebreak based on structural component time series models or more general Kalman filtering (based on the linear case of the general state-space model or hidden Markov model), cf.~\cite{BJR08,EAM95} for detailed description of the corresponding mathematical models.
These approaches can be directly applied to problems of Type~B described in Section~\ref{sec:task} and are based on relating a stochastic model with parameters $\Theta=\{\theta^1,\ldots,\theta^r\}$ to the training part $\{X_t\}_{t=0,1,\ldots,N-1}$ of the observed time series $\{X_t\}_{t}$ so as to make short-term (usually one-step ahead) forecasts, i.e., to estimate the conditional expectation $\mathbb{E}[X_{t+\Delta t}\mid X_{t}, X_{t-1},\ldots; \Theta]$ by $\hat{X}_{t+\Delta t}$ for all $t$ (in particular for $t+\Delta t\geq N$), which basically relies on calculating the maximum likelihood estimate $\hat{\Theta}$ of the parameters $\Theta$, making use of available observations.
Setting a threshold value $\delta$, if the actual observation $X_{t_0}$ varies enough from the forecast value $\hat{X}_{t_0}$ in the sense that $|X_{t_0}-\hat{X}_{t_0}|>\delta$, then the data point $X_{t_0}$ observed at time $t_0$ is considered abnormal.

Among the aforementioned stochastic models, the most demonstrative one is to decompose the underlying time series into trend, seasonal, and independent noise components, where the trend and seasonal components are assumed to be deterministic functions of time which can be fitted by a polynomial and conducting Fourier analysis, respectively.
In fact, this is a special case of the general structural component time series model with trend and seasonal conmponents being stochastic processes.
Each structural component model can be straightforwardly represented as a linear state-space model for which Kalman filtering comes into use to generate forecasts; it also has an equivalent ARIMA model representation for which forecasting can be conducted by following the ARIMA methodology.
The ARIMA approach is based on spectral theory.
For seasonal time series, a parsimonious form termed (multiplicative) seasonal ARIMA (SARIMA) model may be considered.
In general, modelling a time series with an ARIMA representation requires data-specific transformation (i.e.~data pre-processing e.g.~logarithmising, power transformation, and differencing) and a data-adaptive hyperparameter choice (i.e.~the design of the parameter set $\Theta=\{\theta^1,\ldots,\theta^r\}$, in particular the number of parameters $r$) which relies on inspection of the autocorrelogram and partial autocorrelogram.
Each ARIMA model has an equivalent linear state-space model representation allowing Kalman filtering to be employed for forecasting.

The ARIMA approach and Kalman filtering are powerful tools in many applications and in particular in the presence of noise, provided that the hyperparameter choice is reasonable.
However, being the most technically manageble segment of the general state-space models, linear models lack complexity and therefore do not always deliver a feasible approximation for real world applications.
In addition, the associated data-adapted model selection including data pre-processing requires specific expert knowledge and is therefore difficult to implement for fully automatic execution as in our machine-learning framework.
Furthermore, considering problems of Type~A described in Section~\ref{sec:task}, it is unclear how to choose a general representative time series $\{X^*_t\}_t$ in which the diverse variations arising from the individual training signals $\{\{X^{(\iota)}_t\}_t\mid \iota \in I \}$ are incorporated so that the model fitted to $\{X^*_t\}_t$ applies to all normal signals.

\subsubsection{Long Short-Term Memory Units}\label{sec:lstm}
Long short-term memory units (LSTM) are a special type of recurrent neural network (RNN).
As such, the LSTM reads the input time series sequentially, transforming at each point in time the input data into a hidden state which is a nonlinear function of the current input and the hidden state one time step earlier.
The advantage of LSTMs over most other types of RNN is that the dependency of the current on the previous hidden state is designed in such a way that the LSTM obtains the ability to keep (parts of) its hidden state over a larger number of time steps than is possible with other RNN architectures, i.e., LSTMs are able to ``memorise'' values from the past.

Applying LSTMs to prediction tasks for the purpose of anomaly detection works in a similar manner to the application of statistical methods described in the beginning of Section~\ref{sec:ARIMA}.
The main differences are that LSTMs allow for nonlinear parametrisation and have the potential to support a much larger number of parameters which are not estimated directly but instead are randomly initialised at first and then optimised during training (learnt) to obtain the desired predictor.
The complexity of the LSTMs allows them to ingest characteristics of rich and varied training data such as those from large training data sets of Type~A as described in Section~\ref{sec:task} through the process of training with a stochastic gradient descent (SGD) type algorithm.
The training set is processed repeatedly and the parameters of the LSTM are adjusted to optimise the quality of the forecast across the entire training dataset.

Technically, the main drawback of LSTMs is the fact that they are fundamentally still RNNs and hence also suffer from some of the difficulties typical for training this class of artificial neural network such as exploding gradients and a high potential for overfitting.

As a general drawback of using one-step ahead prediction for anomaly detection in time series, if the time series is very complex and exhibits regions in which it is difficult to make precise forecasts, such as when analysing periodic signals containing steep edges or spikes whose positions or values vary randomly over time, reliably estimating the values in these regions can actually be impossible for any type of one-step ahead prediction.
It is thus difficult to derive an anomaly detector from such a predictor as the estimated values can have a large distance to the actual ones and thus show up as false positives.
In Appendix~\ref{sec:comp-lstm} this is illustrated in more detail by training and evaluating an LSTM on the ECG database.

\subsection{Concept of this paper}
Let us now introduce the concept of our machine-learning phase classification approach to the problems specified in Section~\ref{sec:task}.

\subsubsection{Motivation of using convolutional neural networks for phase classification}
Convolutional neural networks (CNNs) are a specific architecture of feed-forward neural networks.
When compared to a fully connected neural network, convolutional neural networks need fewer parameters.
Hence they do not require as large a training dataset and are less prone to overfitting.
CNNs make explicit use of the temporal or spatial structure of the input signal; the signal is analysed locally (local receptive fields) and in a shift invariant manner (translation invariance).
Investiagtions on the internal representations present throughout the layers of CNNs show a high tolerance to noise of various kinds.

Like LSTMs, compared to statistical methods, CNNs have the advantage that, through the use of multiple channels and non-linearities, they provide enough flexibility to capture intricate structures of analysed signals and are able to find representations for large and varied data sets.
They are however easier to train than RNNs, as they suffer less from the vanishing and exploding gradient problems.
The capability of a CNN of being able to process high amounts of complexity has been analysed in the field of image processing, where it was shown \cite{ZZ14} that the neurons inside a convolutional neural network can activate on patterns ranging from simple edges to things as complex as faces.

While CNNs can be used to make forecasts in time series, they particularly excel at classification of spatial or temporal data.
Since the main problem of the LSTM-based approach to anomaly detection in time series outlined above is the general unfeasability of using one-step ahead forecasts, we capitalise on the strength of CNNs in classification tasks and devise a new type of anomaly detection scheme relying on phase classification instead of one-step ahead forecasting.
More details on the properties and operation of CNNs are given in Section~\ref{sec:convnet}.

\subsubsection{Phase classification and anomaly detection}
Motivated by the advantages of convolutional neural networks in classification tasks when dealing with spatial or temporal data, the maching-learning approach proposed in this paper is based on the following key ideas:
\begin{enumerate}
  \item Conducting multi-dimensional time series analysis by means of multi-channel deep convolutional neural networks, where each channel in the input layer corresponds to a single feature (dimension) of the considered time series
  \item Identifying phases or, equivalently, relative locations (order of occurrence) of subpatterns from time series with periodic characteristics by means of training data-adapted classifiers so that subpatterns over different periods of the underlying time series are properly separated into a certain amount of classes
\end{enumerate}

To be more specific, considering a seasonal time series $\{X_t\}_{t}$ with period begins (e.g.~time of local peak values) $\{\tau_k\}_k$, for a pre-determined \emph{initial number of classes} $n_0$, sampling from the original signal $n_0$ overlapping segments per period with a \emph{sliding window of length} $T$, each subpattern $\{X^{(m)}_t\}_{t=0,\ldots,T-1}$ with
\[
X^{(m)}_t:=X_{\tau_{k}+(\tau_{k+1}-\tau_{k}) (m \bmod n_0)/n_0 +t}, \quad k=\lfloor m/n_0\rfloor,
\]
$m\in \mathbb{N}$, is assigned to the class labelled $m \bmod n_0$.

For seasonal data, subpatterns sampled from the time series occur repeatedly and in fixed order within each single period.
A successfully trained classifier outputs the correct class indicating the phase or, equivalently, the relative location in time (i.e.~time distance between subpattern and period begin) of the input subpattern.
Abnormal datapoints in an input pattern are expected to cause false classification results and therefore to be identified as anomalies, which yields a direct solution to problems of Type~B described in Section~\ref{sec:task}.
For problems of Type~A described in Section~\ref{sec:task}, setting a minimum expected classification accuracy (threshold value) and evaluating the classification accuracy of each test signal over a certain number of periods (which is denoted by $K$ in the sequel), those signals that fail to achieve this minimum are considered abnormal.

In order to optimise the classification accuracy of normal data and hence prevent false positive anomaly detection results, we carry out a dynamic reclustering which cancels confusing classes, i.e., subpatterns within a period of the signal that are similar enough to one another are merged into one class.
This reclustering procedure along with the optimisation of the stride length $\Delta t:=s/n_0$ (i.e.~time distance between the segments to be classified) is implemented as a dynamic model selection scheme integrated in our training algorithm.
In addition, we design an auxiliary period detection scheme which is employed in case of a randomly varying period length $s$.

\begin{sidewaysfigure*}[p!]
  \centering
  \includegraphics[width=.85\linewidth]{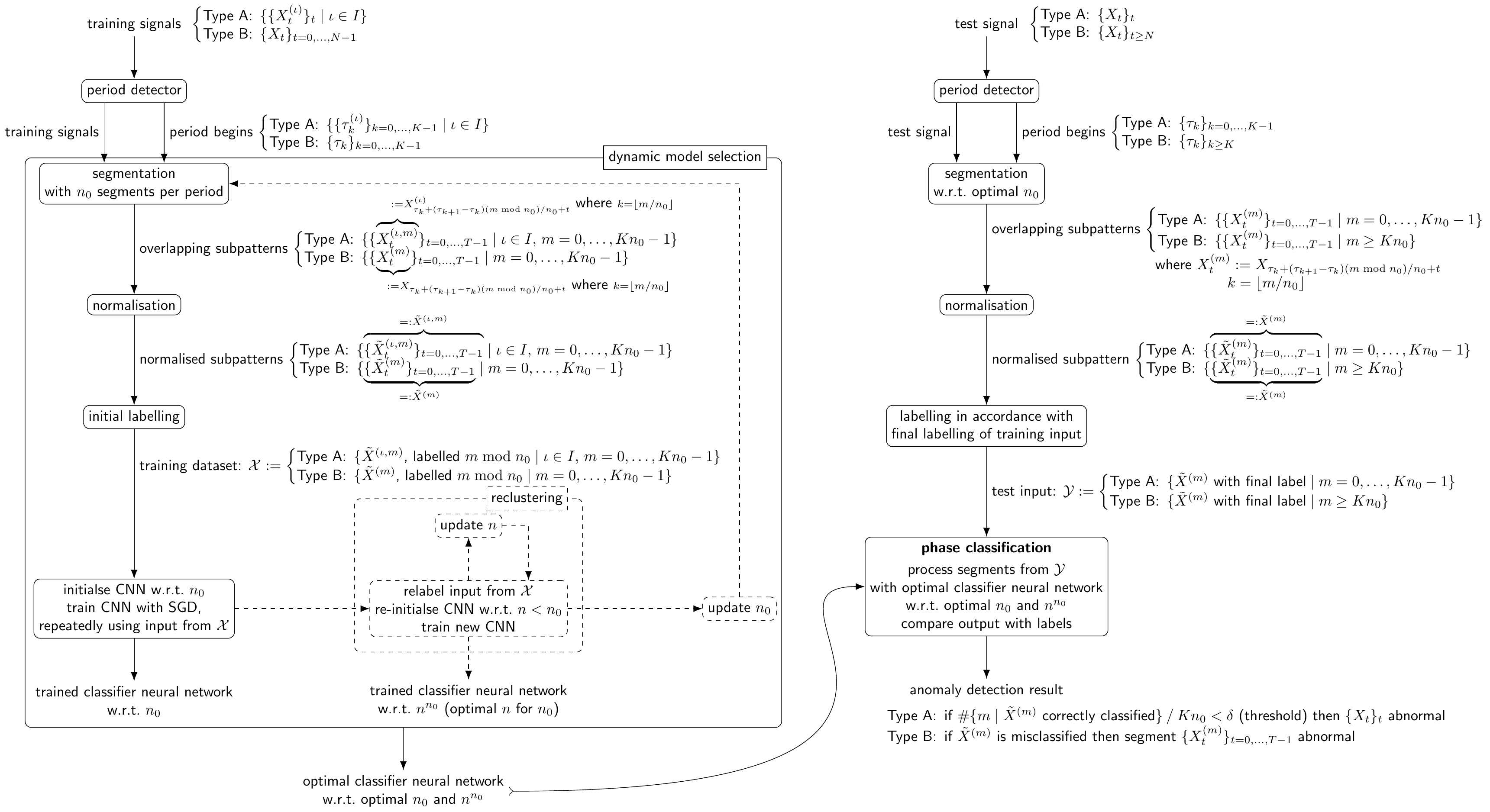}
  \caption{\beginnew Block diagram of training algorithm and anomaly detection\endnew}
  \label{fig:block}
\end{sidewaysfigure*}

The block diagram in Figure~\ref{fig:block} outlines the major steps of our training algorithm and anomaly detection scheme where the steps marked by dashed lines are conditioned by some model-adequacy monitoring criteria which are described in the subsequent section.
\endnew

\section{Method}\label{sec:concept}

\beginnew
In this section we present the general procedure of our phase classification scheme in detail and provide some guidelines for the hyperparameter choice.
\endnew

\subsection{Data pre-processing}\label{sec:pre-processing}
\beginnew
Prior to being fed into the classifier neural nets, all input signals (including training, validation, and test data) are processed by a period detector, cut into overlapping segments by a sliding window, and subsequently normalised, where the segmentation and normalisation depend on the initial number of classes $n_0$.
\endnew

\subsubsection{Period detection}
In general, the seasonal effects of a time series can be recognised by examining the autocorrelogram (cf.~\cite[2.1.4]{BJR08}) or periodogram (cf.~\cite[2.2.1]{BJR08}).
In many cases the period length $s$ is fixed and known.
In case of a fluctuating $s$ (cf.~e.g.~data from cardiology), an auxiliary period detector is designed in Appendix~\ref{pds}, capturing the time of local extremum values (considered as period begins in our setting) $\{\tau_{k}\}_k$ within individual periods and using cross-correlations in order to achieve robust period detection.
Note that in our setting for randomly varying period length $s$, the stride length $\Delta t = s / n_{0}$ while segmenting the signal varies proportionally to $s$ so that the number of overlapping segments from each period is fixed and equal to $n_0$.

\subsubsection{Sliding window}\label{sec:sliding window}
The classification accuracy of our approach turns out not to be highly sensitive to the length of the sliding window $T$.
In the context of anomaly detection, the value of $T$ should be kept relatively small (e.g.~less than or equal to three times the average duration of a single abnormal data sequence) in order to highlight the local effect of the abnormal data points on the time series.
We use a window size of $T=\lfloor 3\overline{s}/n_0\rfloor$ (approximately three times the stride length) where $\overline{s}$ refers to the average value of $s$ (recall that in general $s$ may vary over time).
Empirically, this has proven to be adequate for our purpose.
Note that the length of the sliding window remains constant even in the case of randomly varying period length $s$, the varying stride length merely affects the amount of overlap between adjacent sliding windows.

\subsubsection{Normalisation}\label{subsec:normalise}
In order to remove trend components and avoid skewed results due to dominating extreme values, the samples within the sliding window are normalised by adjusting the local mean and variance, that is, each time considering a $d$-dimensional time series $\{X_t\}_{t}=\{X^i_t\}^{i=0,\ldots d-1}_{t}$ with period begins $\{\tau_k\}_k$ to be processed by a classifier neural network corresponding to initial number of classes $n_0$, for $i=0,\ldots,d-1$ and $m\in \mathbb{N}$, the vector $(\tilde{X}^{i,(m)}_{t})_{t=0,\ldots, T-1}$ is fed into channel $i$ of the convolutional neural net, where
\[
\tilde{X}^{i,(m)}_{t} := \frac{X^{i,(m)}_{t} - \mu^{i,(m)}}{\sigma^{i,(m)}}\quad \text{for } \quad t=0,\ldots,T-1
\]
with
\[
X^{i,(m)}_t:=X^{i}_{\tau_{k}+(\tau_{k+1}-\tau_{k}) (m \bmod n_0)/n_0 +t}, \enskip k=\lfloor m/n_0\rfloor,
\]
\[
\mu^{i,(m)}:=\frac{1}{T}\sum_{t=0}^{T-1}X^{i,(m)}_t,
\]
\[
\sigma^{i,(m)}:=\sqrt{\frac{1}{T}\sum_{t=0}^{T-1}(X^{i,(m)}_t-\mu^{i,(m)})^2}.
\]

\beginnew
For the training and validation data, each subpattern $\tilde{X}^{(m)}:=\{\tilde{X}_t^{(m)}\}_{t=0,\ldots,T-1}$ is initially labelled $m\bmod n_0$.
If reclustering occurs during the training so that the training and validation inputs are relabelled (cf.~Section~\ref{sec:reclustering} for more details), then the test data are labelled in accordance with the final labelling of the training and validation data.
\endnew

\subsection{Convolutional neural networks}\label{sec:convnet}
The core of our phase classifier is a convolutional neural network (CNN).
CNNs are a special type of feedforward neural network, which exploit structures of space or time by sharing many of the weights among different neurons.
We provide a short description of the mathematical basis of a convolutional neural network.
\beginnew
For more detail on the subject, we refer the reader to the literature, e.g.~\cite[Ch.~9]{GBC16}.
\endnew

Basically, a feedforward neural network is a function $f \colon \mathbb{R}^{N^{(0)}} \times \mathbb{R}^{P} \longrightarrow \mathbb{R}^{n}$, mapping an input vector $x \in \mathbb{R}^{N^{(0)}}$ to an output vector $y = f(x, p) \in \mathbb{R}^{n}$, using a vector of parameters $p \in \mathbb{R}^{P}$ to adapt the mapping.
When acting as a classifier, $n$ is the number of classes and the predicted class of a given input $x$ is taken to be $\argmax_{j < n} y_{j}$.
The network can be decomposed into layers, each of which represents a different function mapping vectors to vectors, i.e.,
\[
    f(x, p)
  = f^{(L - 1)}(\cdots f^{(0)}(x, p^{(0)}) \cdots, p^{(L - 1)})
\]
where $L$ is the number of layers and for $l = 0, \dots, L - 1$ the functions $f^{(l)} \colon \mathbb{R}^{N^{(l)}} \times \mathbb{R}^{P^{(l)}} \longrightarrow \mathbb{R}^{N^{(l + 1)}}$ are the transformations performed by each of the single layers and the vectors $p^{(l)} \in \mathbb{R}^{P^{(l)}}$ are again parameter vectors used to adapt the mapping and given as subvectors of $p$.
For ease of notation, let us denote the input to the function $f^{(l)}$ by $x^{(l)}$, starting with $x^{(0)} = x$ and the output of the function $f^{(l)}$ by $x^{(l + 1)}$, ending with $x^{(L)} = y$.

In the most simple feedforward neural networks, each of the transformations $f^{(l)}$ is given by a multiplication with a matrix $a^{(l)} \in \mathbb{R}^{N^{(l)} \times N^{(l + 1)}}$ called the \emph{weight matrix} followed by an addition of a vector $b^{(l)} \in \mathbb{R}^{N^{(l + 1)}}$ called the \emph{bias vector} followed by the application of some non-linear function $g \colon \mathbb{R} \longrightarrow \mathbb{R}$ called the \emph{activation function} to each of the components of the resulting vector, i.e.,
\begin{align*}
         x^{(l + 1)}
     & = f^{(l)}(x^{(l)}, p^{(l)})
  \\ & = g(x^{(l)} \cdot a^{(l)} + b^{(l)})
  \\ & = \biggl( g \biggl( \sum_{i = 0}^{N^{(l)} - 1} x_{i}^{(l)} \cdot a_{i, j}^{(l)} + b_{j}^{(l)} \biggr) \biggr)_{j < N^{(l + 1)}}
  .
\end{align*}
The entries of the matrix $a^{(l)}$ and the vector $b^{(l)}$ are exactly the components of the parameter vector $p^{(l)}$.
This is called a \emph{fully connected} layer.

In the case of a one-dimensional \emph{convolutional layer}, the affine transformation $x^{(l)} \longmapsto x^{(l)} \cdot a^{(l)} + b^{(l)}$ is replaced with a more restrictive kind of affine transformation, the so-called batched convolution.
For this, the vector $x^{(l)} = (x_{i}^{(l)})_{i < N^{(l)}}$ is reindexed to form a two-dimensional array $(x^{(l)}_{i, t})_{i < M^{(l)}, t < T^{(l)}}$ with $M^{(l)} \cdot T^{(l)} = N^{(l)}$ and we say that the sequence $(x^{(l)}_{i,t})_{t<T^{(l)}}$ is fed into the $i$-th \emph{channel} of the convolutional layer $l$.%
\footnote{%
\beginnew
Vice versa, when placing a fully connected layer after a convolutional layer, the inverse reindexing is performed.
When using a convolutional layer as the first layer of an artificial neural network and the input is in fact a segment of a multivariate time series $\{ X_{i, t} \}_{i < M, t}$ with $M = M^{(0)}$ features, no reindexing is required.
This is the case in all of our set-ups.%
\endnew
}
Similarly, the parameter vector $p^{(l)}$ is distributed not into a matrix $a^{(l)}$ and a vector $b^{(l)}$, but into a matrix of vectors $(k_{i, j, s}^{(l)})_{\substack{i < M^{(l)}, j < M^{(l + 1)}, s < S^{(l)}}}$ called the \emph{convolution kernels} and a vector $b^{(l)} \in \mathbb{R}^{M^{(l + 1)}}$.
The operation performed by the function $f^{(l)}$ is now given by
\begin{align*}
     & \phantom{{} = {}}
         x^{(l + 1)}
  \\ & = f^{(l)}(x^{(l)}, p^{(l)})
  \\ & = g(x^{(l)} \mathbin{\dot{*}} k^{(l)} + b^{(l)})
  \\ & = \biggl( g \biggl( \sum_{\substack{i < M^{(l)} \\ s < S^{(l)}} \hspace{-.5em}} x_{i, t + S^{(l)} - 1 - s}^{(l)} \cdot k_{i, j, s}^{(l)} + b_{j}^{(l)} \biggr) \biggr)_{\substack{j < M^{(l + 1)} \\ t < T^{(l + 1)}}}
\end{align*}
and we have the constraint that $T^{(l + 1)} = T^{(l)} - S^{(l)} + 1$.
In many convolutional networks, the input vectors $(x^{(l)}_{i, t})_{t < T^{(l)}}$ are extended (padded) by additional zero entries prior to being convolved.
When padding with exactly $S^{(l)} - 1$ zeros, the output vectors are of the same size as the input vectors.
\beginnew
If furthermore the padding is performed symmetrically, i.e., if $(S^{(l)} - 1) / 2$ zeros are added to both ends of the signal, this is referred to as `SAME'-padding.
\endnew

We also use a type of layer called a \emph{max pooling layer} between two convolutional layers.
The transfer function $f^{(l)} \colon \mathbb{R}^{M^{(l)} \times T^{(l)}} \longrightarrow \mathbb{R}^{M^{(l + 1)} \times T^{(l + 1)}}$ of this layer is given by
\[
    f^{(l)}(x^{(l)})
  = \Bigl( \max_{\substack{r < R^{(l)} \\ t \cdot R^{(l)} + r < T^{(l)}}} x_{j, t \cdot R^{(l)} + r}^{(l)} \Bigr)_{\substack{j < M^{(l + 1)} \\ t < T^{(l + 1)}}}
\]
where $R^{(l)}$ is a positive integer called the \emph{pool size} and we have the constraints $M^{(l + 1)} = M^{(l)}$ and $T^{(l + 1)} = \lceil T^{(l)} / R^{(l)} \rceil$.
Note that max pooling layers have no adjustable parameters $p^{(l)}$.

In our networks, we employ both convolutional and regular fully connected layers.
We apply `SAME'-padding in all convolutional layers and use the hyperbolic tangent ($\tanh$) as activation function $g$ throughout the entire network, which is a common choice in feedforward neural networks.

The exact layout of the convolutional network used for our task is displayed in Table~\ref{tab:net-layout}.
\begin{table}[htbp!]
  \caption{Layers of classifier neural network}
  \label{tab:net-layout}
  \centering
  \begin{tabular}{lll}
       \hline
         & \textbf{Layer Type} & \textbf{Sizes}
    \\ \hline
       0 & convolutional   & $M^{(0)} = d$, $T^{(0)} = T$,
    \\   &                 & $S^{(0)} = (\lfloor T / 6 \rfloor + 1) \cdot 2 + 1$
    \\ 1 & max pooling     & $M^{(1)} = M^{(0)} \cdot 6$, $T^{(1)} = T^{(0)}$,
    \\   &                 & $R^{(1)} = 3$ if applicable, else $R^{(1)}=1$
    \\ 2 & convolutional   & $M^{(2)} = M^{(1)}$, $T^{(2)} = \lceil T^{(1)} / R^{(1)} \rceil$,
    \\   &                 & $S^{(2)} = (\lfloor T / 12 \rfloor + 1) \cdot 2 + 1$
    \\   &                 & $M^{(3)} = M^{(2)} \cdot 3$, $T^{(3)} = T^{(2)}$
    \\ 3 & fully connected & $N^{(3)} = M^{(3)} \cdot T^{(3)}$
    \\ 4 & fully connected & $N^{(4)} = \lfloor \sqrt{N^{(3)} \cdot N^{(5)}} \rfloor$
    \\ 5 & output          & $N^{(5)} = n$
    \\ \hline
  \end{tabular}
\end{table}
Here $d$, $T$, and $n$ denote the dimension of the input time series, the sliding window size, and the current number of classes, respectively.
The layer and kernel sizes are chosen to best adapt to varying input time series dimensions, sliding window sizes, and numbers of classes.
\beginnew
In the convolutional layers, the number of channels is increased first by a factor of 6, then by a factor of 3.
Such increases are common in convolutional neural networks and allow the layers to capture different aspects of the incoming signal such as edges and more complex patterns.
\endnew

The method, by which the parameter vector $p$ is adjusted and the network adapts, is the minimisation of a function $h \colon \mathbb{R}^{n} \longrightarrow \mathbb{R}$ applied to the output of the neural network, called the \emph{loss function}.
Since we are classifying phases, to each training input $x$ (and hence to each output $y$), there corresponds a label $z \in \{ 0, \ldots, n - 1 \}$.
In our case, we use the cross entropy loss function, which is given by
\[
    h(y, z)
  = -w_{z} \log \biggl( \frac{\exp y_{z}}{\sum_{j = 0}^{M - 1} \exp y_{j}} \biggr)
\]
where $w_{z}$ denotes a weight by which the losses of each class are scaled.
The weights are statically determined and are in our case chosen to be proportional to the inverse of the number of training examples for each class in order to counteract bias caused by unbalanced classes.

\subsection{Training algorithm}\label{subsec:conv net and classifier}
The neural networks in our algorithm are trained by the ADAM training algorithm which is a refined version of stochastic gradient descent (SGD).
In SGD, the average loss for a set $\mathcal{X}_{\operatorname{batch}}$ containing pairs of training inputs $x$ and corresponding labels $z$ is minimised by changing the randomly initialised parameters $p$ of the neural network according to the update rule
\[
        p
  \gets p - \frac{\gamma}{\# \mathcal{X}_{\operatorname{batch}}} \sum_{(x, z) \in \mathcal{X}_{\operatorname{batch}}} \nabla_{p} h(f(x, p), z)
\]
where $\gamma$ is a tuning hyperparameter called the \emph{learning rate} and $\nabla_{p}$ is the gradient operator with respect to the vector of parameters $p$.
This minimises the average of the loss values $h(f(x, p), z)$.
\beginnew
The gradient $\nabla_{p} h(f(x, p), z)$ is computed in an efficient manner via reverse-mode auto differentiation which is basically an application of the chain rule.
This is also known as the backpropagation algorithm and more details on the process can be found in the literature, e.g.~\cite[Ch.~4]{GBC16}.
\endnew
The set $\mathcal{X}_{\operatorname{batch}}$ is called a $\emph{mini-batch}$ and is taken to be a subset of the set of all available training inputs $\mathcal{X}$.
The update steps are performed with changing disjoint mini-batches until the entire training dataset $\mathcal{X}$ is exhausted.
Each pass through the entire set of available training data is referred to as an \emph{epoch}.
To enhance the training process (cf.~\cite{SKL18}), for rich datasets we change the size of the mini-batches during training, later epochs use larger mini-batch sizes.
The adaptive adjustments performed by the ADAM algorithm detailed in \cite{KB14} provide further enhancements to this process.

In contrast to usual classifiers, our algorithm encapsulates the gradient descent algorithm in a decision process monitoring the necessity of \emph{dynamic reclustering} which aims to optimise the classification accuracy.
\beginnew
The complete algorithm is given in Algorithm~\ref{alg:training} (cf.~also Figure~\ref{fig:block}), the single steps are described in more detail in the remainder of this section.
\endnew
\begin{algorithm}[htbp!]
  \caption{Training algorithm}
  \label{alg:training}
  \begin{algorithmic}
    \State $n_{\operatorname{best}} \gets 0$
    \For{$n_{0} \in \{ n_{0}^{*}, n_{0}^{*} - 2, \ldots, 4 \}$}
      \If{$n_{0} \le n_{\operatorname{best}}$}
        \State\Return stored net
      \EndIf
      \State $n \gets n_{0}$
      \While{\textbf{true}}
        \State initialise net and labels
        \Repeat
          \State perform training iteration
        \Until{no improvement in validation loss\\\hfill within $4$ consecutive epochs}
        \State \Comment{training stop crit.}
        \If{minimum class accuracy $\ge 1 - \alpha$}
          \State \Comment{reclustering stop crit.}
          \State store net
          \State $n_{\operatorname{best}} \gets n$
          \State \textbf{break}
        \EndIf
        \If{$n - 1 < 3$ \textbf{or} $n - 1 \le n_{\operatorname{best}}$}
          \State \textbf{break}
        \EndIf
        \State $n \gets n - 1$
        \State recluster according to overall confusion matrix
        \State update weights of loss function
      \EndWhile
    \EndFor
    \If{$n_{\operatorname{best}} \ne 0$}
      \State\Return stored net
    \Else
      \State change $\alpha$ and rerun
    \EndIf
  \end{algorithmic}
\end{algorithm}

Each time having initialised the neural network for separating the currently considered classes, the gradient descent optimiser is run until a training-progress-monitoring stop criterion is fulfilled (cf.~\emph{training stop criterion} in Section~\ref{sec:stop criteria} for more details).
The classification ability of the underlying neural net is evaluated by means of the so-called \emph{confusion matrices} (cf.~Section~\ref{sec:confusion}) throughout the entire training.
If at the end of training all classes are evaluated with sufficient accuracy (cf.~\emph{reclustering stop criterion} in Section~\ref{sec:stop criteria}), the trained neural net is stored; otherwise a relabelling procedure according to the \emph{overall confusion matrix} is conducted where the class with least average evaluation accuracy is merged into the class to which the corresponding inputs are most commonly misclassified during training and the neural net is re-initialised with respect to the updated classes (cf.~Section~\ref{sec:reclustering}).
Among all stored neural nets, the ultimate classifier is chosen as the one having the maximum number of output classes (cf.~Section~\ref{sec:final number classes}).

In the subsequent subsections, the aforementioned reclustering process and stop criteria are described in detail.

\subsubsection{Confusion matrix}\label{sec:confusion}
In order to track the progression of classification accuracy during training, we record the \emph{confusion matrix} evaluated on the \emph{training data} after each epoch.
For a current number of classes $n$ and existing classes labelled as $0,\dots,n-1$, the \emph{confusion matrix} evaluated after the $k$-th epoch is an $(n\times n)$-dimensional matrix denoted by $V_k^{(n)}:=(V_k^{ij,(n)})_{i,j=0,1,\dots,n-1}$, where the entry $V_k^{ij,(n)}$ refers to the number of training inputs labelled as $i$ and predicted by the neural net during the $k$-th training epoch as class $j$, $k \ge 0$.

During the experimentation, we observe that classes which are easily distinguishable can already be separated after very few training iterations, whereas classes sharing more similarity perform significantly worse in the beginning and also show a slower increase of evaluation accuracy during training.
Taking into account that the evaluated value of the loss function commonly follows a convex decreasing trend throughout the entire training, the above observation motivates us to assess the separation ability of the underlying neural net during training by weighting the confusion matrix with the respective contribution to the training progress and to introduce the \emph{overall confusion matrix} denoted by $\overline{V}^{(n)}:=(\overline{V}^{ij,(n)})_{i,j = 0,\dots,n-1}$ and defined as
\begin{equation}\label{eq:overall confusion}
\overline{V}^{ij,(n)}:= \sum_{k=1}^{E^{(n)}-1} V_{k}^{ij,(n)}(H_{k-1} ^{(n)}- H_{k}^{(n)}),
\end{equation}
where $n$, $E^{(n)}$, and $H_k^{(n)}$ refer to the current number of classes, the number of training epochs that are performed until the training stop criterion (cf.~Section~\ref{sec:stop criteria}) is satisfied, and the average training loss during the $k$-th epoch, respectively.

In our setting, the confusion matrices serve as the key objects of the decision criteria for our dynamic reclustering (cf.~Sections~\ref{sec:stop criteria} and~\ref{sec:reclustering}).
The definition of the overall confusion matrix in terms of \eqref{eq:overall confusion} by taking the weighted average throughout the entire training and dropping the values from the initial epoch ($k = 0$) aims to mitigate the random effect of the initialisation of the neural network.
Empirically, this yields robust reclustering results during different test runs for fixed $n_{0}$.

\subsubsection{Stop criteria}\label{sec:stop criteria}
The criteria for stopping the loops are related to parametrised effectiveness and accuracy reqirements in the following manner:

\begin{description}
  \item[Training stop criterion]
    We monitor the training progress by evaluating the average loss of validation data over each training epoch.
    For each (re-)initialised neural network, training is stopped if no improvement in the average validation loss during the latest $4$ epochs can be observed.

  \item[Reclustering stop criterion]
    Allowing a \emph{maximum per-class margin of error} $\alpha\in[0,1)$, the reclustering procedure is stopped if
    \[
    \min_{i=0,\dots,n-1}\frac{V_{E^{(n)}-1}^{ii,(n)}}{\sum_{j=0}^{n-1}V_{E^{(n)}-1}^{ij,(n)}}\geq 1-\alpha,
    \]
    where $n$, $E^{(n)}$, and $V_{E^{(n)}-1}^{(n)}$ refer to the current number of classes, number of epochs for training the related network (i.e.~until the training stop criterion is fulfilled), and the respective confusion matrix evaluated at the end of training (recall the definition in Section~\ref{sec:confusion}), respectively.
\end{description}

\beginnew
By definition of the confusion matrix in Section~\ref{sec:confusion}, for $k=E^{(n)}-1$ and each $i=0,\ldots,n-1$, the diagonal element $V^{ii,(n)}_{k}$ divided by the respective row sum $\sum_{j=0}^{n-1} V_k^{ij,(n)}$ of the confusion matrix is the share of correctly classified training inputs in all training inputs labelled as $i$ evaluated during the last epoch while training the classifier neural network with $n$ existing classes.
Therefore, for a pre-defined margin of error $\alpha\in[0,1)$, the above reclustering stop criterion requires that at the end of training the corresponding classifier neural network should correctly classify the training inputs of each existing class at least at the rate of $1-\alpha$.
\endnew

\subsubsection{Reclustering}\label{sec:reclustering}
As long as the recustering stop criterion is not fulfilled, the subsequent reclustering procedure is considered necessary.

For a current number of classes $n$ and existing classes labelled as $0,\dots,n-1$, let $i^\circ$ and $j^{\circ}$ denote the worst evaluated class and the correponding most misassigned class during the entire training of the respective neural net (i.e.~until the training stop criterion is fulfilled) which are defined as
\[
i^\circ:= \argmin_{i=0,\dots,n-1} \frac{\overline{V}^{ii,(n)}}{\sum_{j=0}^{n-1}\overline{V}^{ij,(n)}}
\]
and
\[
j^{\circ}:= \argmax_{\substack{j=0,\dots,n-1\\j\neq i^{\circ}}} \overline{V}^{i^{\circ} j,(n)}
\]
respectively (recall the definition of $\overline{V}^{(n)}$ in \eqref{eq:overall confusion}).
The class labelled as $i^\circ$ is merged into class $j^{\circ}$.
Furthermore, since we always assume the labels to be consecutive, the training and validation inputs with the largest label $n-1$ are assigned the label of the dropped class $i^\circ$.

Each time after relabelling, the weights corresponding to the remaining classes in the cost function are adjusted to be again inversely proportional to the current shares of the classes in order to warrant a well-balanced training of the updated classifier and the neural net is re-initialised.

\subsubsection{Final number of classes}\label{sec:final number classes}
In the context of anomaly detection, we are dealing with the trade-off between optimising the classification accuracy of normal data preventing false positives (i.e.~to cancel confusing classes) and maintaining the ability of misclassifying abnormal data for the sake of anomaly detection (i.e.~to still retain sufficiently many classes characterising different phases within a period).
Keeping this in mind, the final number of classes determining the ultimate classifier neural network is selected in the following manner:

Given a maximum allowed number of classes $n_0^*$ with $n_0^*$ an even number $n_0^*>3$, the starting initial number of classes is set to $n_0:=n_0^*$.
Each time for an updated initial number of classes $n_0$, the relabelling procedure described in Section~\ref{sec:reclustering} is run at most $(n_0-3)$-times (i.e.~with at least $3$ remaining classes).
If the reclustering stop criterion is fulfilled after relabelling $\Delta n^{n_0}$-times, the \emph{candidate final number of classes related to $n_0$} is set to $n^{n_0}:=n_0-\Delta n^{n_0}$ and the corresponding neural net is stored.
If $\max_{n'_0 = n^*_0, \ldots, n_{0}} n^{n'_0} \ge n_0 - 2$, the updating processes of $n_0$ is finished; otherwise $n_0$ is reduced by $2$.
The \emph{overall final number of classes} refers to the maximum of $n^{n_0}$ taken along the entire path of $n_0$, i.e.~$\max_{n'_0 = n^*_0, \ldots, 4} n^{n'_0}$ and the final classifier neural network is the one stored when this overall maximum was achieved.
If this maximum was achieved more than once, we choose the neural network corresponding to the highest $n_0$ such that $n^{n_0}$ achieved this maximum.
This is because a high value of $n_0$ corresponds to a narrow sliding window (cf.~Section~\ref{sec:sliding window}) and hence maximises the sensitivity of the anomaly detector.

If in the end no suitable network has been stored, we increase $\alpha$ and rerun the algorithm.
\\[1em]
Finally, it is worth mentioning that once all the hyperparameters are determined, the whole training algorithm introduced above, including data pre-processing and dynamic reclustering, is implemented in a machine-learning manner so that the classification and anomaly detection process can be accomplished fully automatically.

\beginnew
\subsection{Anomaly detection}\label{sec:anom-detect}
Once training is finished and in particular when the ultimate classifier neural network determined by the model selection process turns out to use initial number of classes $n_0$ and final number of classes $n^{n_0}$, each test signal is pre-processed following the procedure described in Section~\ref{sec:pre-processing} with respect to $n_0$, labelled with respect to $n^{n_0}$ in accordance with the training and validation data (recall the relabelling step along with the dynamic reclustering described in Section~\ref{sec:reclustering}), and then processed by the trained ultimate classifier neural network.

For problems of Type~A described in Section~\ref{sec:task}, a minimum expected per-signal average classification accuracy $\delta$ (threshold value) should be set depending on individual needs.
For instance, $\delta$ could be determined on the basis of classification accuracy on validation data.
For each test signal $\{X_t\}_t$ recorded over $K$ periods of time with period begins $\{\tau_k\}_{k=0,\ldots,K-1}$, if the normalised segments $\tilde{X}^{(m)}$, $m=0,\ldots,Kn_0-1$ (recall Section~\ref{subsec:normalise}), processed by the ultimate classifier neural net are evaluated with an average classification accuracy rate less than $\delta$, i.e., if $\#\{m\mid \tilde{X}^{(m)} \text{ correctly classified}\}/Kn_0 < \delta$, then the signal $\{X_t\}_t$ is considered abnormal.

Considering problems of Type~B described in Section~\ref{sec:task}, if a normalised segment $\{\tilde{X}_t^{(m)}\}_{t=0,\ldots,T-1}$ (recall Section~\ref{subsec:normalise}) of the test signal $\{X_t\}_{t\ge N}$ is misclassified by the ultimate classifier neural net, then the original segment $\{X^{(m)}_t\}_{t=0,\ldots,T-1}$ with
\[
X^{(m)}_t=X_{\tau_{k}+(\tau_{k+1}-\tau_{k}) (m \bmod n_0)/n_0 +t}, \quad k=\lfloor m/n_0\rfloor
\]
is considered abnormal.
\endnew

\section{Experiments}\label{sec:example datasets}

In this section we apply our machine-learning algorithm proposed in Section~\ref{sec:concept} to three example datasets choosing from the domains of cardiology, industry, and signal processing, aiming to show the feasibility of the method in a range of applications.
The cardiology dataset is the most complex and challenging dataset representing problems of Type~A described in Section~\ref{sec:task}, as the recordings taken from healthy control patients exhibit a high level of diversity which needs to be captured by the classifier.
This diversity mandates the use of a more complex representation which is one of the strengths of deep neural networks over other parametric models.
The other two datasets demonstrate the applicability of the method in different contexts, including the detection of anomalies occurring only at certain instances in time and thus representing problems of Type~B described in Section~\ref{sec:task}.

\subsection{Cardiology dataset}\label{sec:cardio}
The PTB Diagnostic ECG Database is a database created by the Physikalisch-Technische Bundesanstalt (PTB) consisting of 549 electrocardiogram (ECG) recordings gathered from 290 subjects aged 18 to 87.
The ECGs were recorded using a non-commercial PTB prototype recorder, the specifications of which can be found on the database website\footnote{\url{https://physionet.org/physiobank/database/ptbdb/}}.
The dataset is part of PhysioNet~\cite{GAG00} and is further described in~\cite{BKS95}.

\subsubsection{Input data} \label{sec:cardio-input}
We use $3/5$ and $1/5$ of the measurements from healthy patients for training and validation, respectively.
The trained classifier is tested on the remainder of the data from healthy patients and data from all ill patients.

Due to the large data volume, we manually resample the input data to a sample rate of $50$ samples per second instead of the original $1000$ before feeding it into the neural network (i.e.~the actual time unit applied in our training amounts to $1 \text{ time unit} = 20 \, \text{ms}$).
This operation is not strictly necessary, but it speeds up the training process.
Also, we only use the first 60 periods of each recording during training and for testing.
We train our classifier with resampled time series from healthy patients and use the data coming from all $12$ conventional leads and $3$ Frank leads (cf.~\cite{Fra56}) for the ECG diagnostic, resulting in a convolutional neural net with $15$ channels on the input layer.

\subsubsection{Period detection}\label{sec: peak detector}
The first challenge when analysing ECG data consists in detecting the randomly varying periods of individual patients, for which we design a period detector.
This detector is described in greater detail in Appendix~\ref{pds}.
The detector has a number of parameters which need to be adjusted to the dataset, the actual values used here are given in Table~\ref{tab:ecg-pds-params}.
\begin{table}[htbp!]
  \caption{Parameters for period detector on ECG database}
  \label{tab:ecg-pds-params}
  \centering
  \begin{tabular}{lr}
       \hline
       \textbf{Parameter}                              & \textbf{Value}
    \\ \hline
       prefilter window half-length $n$                &   $10$
    \\ minimum base period length $s_{\min}$           &  $500$
    \\ maximum base period length $s_{\max}$           & $2000$
    \\ maximum period length deviation factor $\sigma$ &  $1/2$
    \\ reference window half-length factor $\lambda$   &  $1/3$
    \\ \hline
  \end{tabular}
\end{table}
For this dataset, the entire time series for feature `i' is used as both the reference and input time series to the period detector.
However, in order to ensure the requirement that no trend component exists in the signal, the first difference of the signal is used instead of the raw signal.
In order to adjust for the offsets thus introduced at peak detection, between Steps~\ref{pds-refsegment} and~\ref{pds-xcorr} described in Appendix~\ref{pds}, the reference window is adjusted to be precisely centred on the corresponding peak in the original (smoothed but not differentiated) signal, i.e., its midpoint $T_{k_{0}}$ is changed to
\[
  \argmax_{T_{k_{0}} - 10 \le t \le T_{k_{0}} + 10} X_{t}
  .
\]
The maximum allowed adjustment of $10$ has empirically been found to yield satisfactory results.

The median of all observed period lengths approximately amounts to $\overline{s}=700 \, \text{ms}=35\text{ time units}$.

\beginnew
\subsubsection{Hyperparameters}
During the training, the maximum allowed number of classes and per-class margin of error are set to $n_0^*:=10$ and $\alpha:= 2^{-5}$, respectively.

As per description in Table~\ref{tab:net-layout}, each of the classfier neural networks encountered during the run consists of two convolutional layers with $M^{(0)}=15$ and $M^{(1)}=90$ channels, respectively, with max pooling of size $R^{(1)}=3$ applied in between, followed by two fully connected layers, and the output layer.
During the classifier selection process, the length $T^{(0)}$ of the input sequence, the kernel sizes $S^{(0)}, S^{(2)}$ of the convolutional layers, and the size $N^{(3)}$ of the first fully connected layer vary proportionally to the sliding window length $T=\lfloor3\overline{s}/n_0\rfloor=\lfloor 105/n_0\rfloor$ where $n_0$ runs over the values in $\{10(=n_0^*),8,6,4\}$ if not stopped earlier.
The size $N^{(4)}$ of the second fully connected layer is determined by the geometric mean of the sizes $N^{(3)}, N^{(5)}$ of its adjacent layers and the size $N^{(5)}$ of the output layer is equal to the current number of classes $n$ which runs over the values in $\{n_0,n_0-1,\ldots\}$ during the dynamic reclustering.

The ADAM optimiser with learning rate $\gamma=0.1$ is employed for training with SGD.
We start at a mini-batch size of $800$ and increase it after every $2$ or $3$ epochs up to $4800$.
\endnew

\subsection{SCADA dataset}\label{sec:SCADA}
In~\cite{LF16}, Antoine Lemay and Jos\'{e} M.\ Fernandez describe a simulation of an industrial control system, specifically designed for providing supervisory control and data acquisition (SCADA) network datasets for intrusion detection research.
The generated datasets are openly available on GitHub\footnote{\url{https://github.com/antoine-lemay/Modbus_dataset}} and contain periods of regular operation, manual interactions with the system, and anomalies caused by network intrusions.
Since the operation of the simulated system is cyclic, the resulting data is mostly periodic.

\subsubsection{Input data}
Among the available datasets with common characteristics, we choose the first $4/5$ and the last $1/5$ of the dataset named `characterization\_\allowbreak modbus\_\allowbreak 6RTU\_\allowbreak with\_\allowbreak operate' with a duration of 5.5 minutes in total for training and validation, respectively, where neither the injected malicious activities nor the manual operations included are labelled, both resulting in a certain proportion of noise in the corresponding time series.
The trained classifier is tested on the only three correctly labelled datasets `moving\_\allowbreak two\_\allowbreak files\_\allowbreak modbus\_\allowbreak 6RTU' (`Test Data 1'), `CnC\_\allowbreak uploading\_\allowbreak exe\_\allowbreak modbus\_\allowbreak 6RTU\_\allowbreak with\_\allowbreak operate' (`Test Data 2'), and `send\_\allowbreak a\_\allowbreak fake\_\allowbreak command\_\allowbreak modbus\_\allowbreak 6RTU\_\allowbreak with\_\allowbreak operate' (`Test Data 3'), including no manual operations, a small portion of manual operations, and a large amount of noise e.g.~manual operations (causing non-intrusion anomalies), respectively.
In each dataset, four features are considered: number and total size of sent packets, and number of active IP address and port pairs.
At one-second intervals, we record the increase in each feature and consider the corresponding $4$-dimensional time series.

The given 10-seconds polling interval yields periodic characteristics of the considered time series with a fixed period length of $s=10 \text{ seconds}$.

\beginnew
\subsubsection{Hyperparameters}
We set $\alpha:= 2^{-3}$ and $n^*_0:= 10$ for training the classifier neural networks.

According to Table~\ref{tab:net-layout}, all convolutional neural networks considered during the entire run include $M^{(0)}=4$ and $M^{(1)}=24$ channels on the first and second convolutional layers, respectively, and two fully connected layers placed between the last convolutional layer and the output layer.
Considering the short input sequence length $T^{(0)}=\lfloor 3s/n_0\rfloor=\lfloor 30/n_0\rfloor$ with $n_0$ taking values in $\{10(=n_0^*),8,\ldots\}$, we do not apply any max pooling, i.e.~$R^{(1)}=1$.
During the classifier selection process, the sizes $S^{(0)}$, $S^{(2)}$ and $N^{(3)}$ of the connvolution kernels and the first fully connected layer, respectively, vary proportionally to the input length $T^{(0)}$.
The size of the output layer $N^{(5)}$ is equal to the current number of classes $n$ which runs over the values in $\{n_0,n_0-1,\ldots\}$ during the dynamic reclustering and the size of its preceding fully connected layer $N^{(4)}$ is the geometric mean of $N^{(5)}$ and $N^{(3)}$.

The ADAM optimiser with learning rate $\gamma=0.01$ and a mini-batch size of $4$ are used for training with SGD.
\endnew

\subsection{Wave dataset}\label{sec: waves}
The waves dataset is a synthetic dataset loosely modelled on a system transmitting a periodic signal.
From the theory of Fourier analysis, every differentiable periodic signal $\{ x_{t} \}_{t}$ with frequency $f$ can be decomposed into its frequency components
\[
    x_{t}
  = a_{0} + \sum_{k = 1}^{\infty} a_{k} \cos(2 \pi (f k t + \varphi_{k}))
  ,
\]
cf.~\cite[Theorem~2.1]{SS03}, which motivates the principal rule of our wave generator.
In our consideration, the generated waves have no DC offset, i.e.~$a_{0} := 0$, and components only up to frequency $4 f$, i.e.~$a_{k} := 0$ for all $k \ge 5$.
The signals are supposed to be transmitted over a noisy channel which we assume to add filtered Brownian and white noise.
The wave generator also has some inherent randomness in the form of clock jitter, amplitude noise, and phase noise.
There are also a number of fault conditions which form the basis of the anomalies to be detected.

\subsubsection{Wave generator}
\label{sec:wave-gen}
The waves in this dataset are of the form
\[
    X_{t}
  = \sum_{k = 1}^{4} R_{t}^{\operatorname{amp} k} \cos(2 \pi (f k T_{t} + R_{t}^{\operatorname{ph} k})) + R_{t}^{\operatorname{noise}} + N_{t}
  ,
\]
$t = 0, 1, 2, \dots$, with $T_{t}$ given by
\[
    T_{t}
  = \sum_{u = 0}^{t - 1} R_{u}^{\operatorname{time}}
  \quad \text{for } t = 0, 1, 2, \dots
\]
and $f := 2^{-8}$.
Here, $\{ N_{t} \}_{t}$ is a Gaussian white noise process, i.e., $N_{0}, N_{1}, N_{2}, \dots$ are independent and identically distributed (i.i.d.) random variables with $N_{t} \sim \mathcal{N}(0, \sigma^{2})$ for all $t = 0, 1, 2, \dots$, and $\{ R_{t}^{\operatorname{amp} k} \}_{t}$, $\{ R_{t}^{\operatorname{ph} k} \}_{t}$, $\{ R_{t}^{\operatorname{noise}} \}_{t}$, and $\{ R_{t}^{\operatorname{time}} \}_{t}$ are independent (discrete) Ornstein-Uhlenbeck processes with individual sets of parameters.
In general, an Ornstein-Uhlenbeck process $\{ R_{t} \}_{t}$ obeys the stochastic differential equation
\begin{equation}
    d R_{t}
  = \theta (\mu - R_{t}) \, d t + \sigma \, d W_{t}
  ,
  \label{equ:ou-sde}
\end{equation}
where $\mu \in \mathbb{R}$, $\sigma > 0$, $\theta \in [0, 1]$, and $\{ W_{t} \}_{t}$ is a standard Brownian motion, cf.~e.g.~\cite[Ex.\,6.6]{Shr04}.
In discrete time, a process $\{ R_{t} \}_{t = 0, 1, 2, \dots}$ following \eqref{equ:ou-sde} can be approximated by generating i.i.d.\ random variables $\tilde{N}_{0}, \tilde{N}_{1}, \tilde{N}_{2}, \dots$ with $\tilde{N}_{t} \sim \mathcal{N}(\mu, (\sigma / \theta)^2)$ for all $t = 0, 1, 2, \dots$ and exponentially smoothing them:
\begin{equation}
     R_{t + 1}
  := \theta \tilde{N}_{t} + (1 - \theta) R_{t}
  \quad \text{for } t = 0, 1, 2, \dots.
  \label{equ:ou-exp}
\end{equation}
Indeed, letting
\[
     N_{t}^{*}
  := \frac{\tilde{N}_{t} - \mu}{\sigma / \theta}
  \quad \text{for } t = 0, 1, 2, \dots,
\]
the process $\{ W_{t} \}_{t=0,1,2,\dots}$ with
\[
     W_{t}
  := \sum_{u = 0}^{t - 1} N_{u}^{*}
\]
is a random walk with Gaussian increments and thus corresponds to a discretely sampled standard Brownian motion~\cite[(1.9)]{RY99}.
Therefore, \eqref{equ:ou-exp} can be written as
\begin{align*}
         R_{t + 1} - R_{t}
     & = \theta \biggl( \mu + \frac{\sigma}{\theta} N_{t}^{*} \biggr) - \theta R_{t}
  \\ & = \theta (\mu - R_{t}) + \sigma (W_{t + 1} - W_{t})
  ,
\end{align*}
which yields a discrete counterpart of \eqref{equ:ou-sde}.
The Ornstein-Uhlenbeck process can be thought of as a process performing a random walk where the increments are biased towards the mean $\mu$.
As such, it behaves locally like a Brownian motion, causing the power of the higher frequency parts of its spectrum to average $1 / f^{2}$ (brownian noise).
The process can be used to model parameters of systems that tend to shift over time, while generally remaining close to a certain average value.

For each single wave, a set of parameters controlling the governing processes is randomly generated using the parameters in Table~\ref{tab:wave-gen-params}.
\begin{table}[htbp!]
  \caption{Parameters for processes governing generated waves ($k = 1, 2, 3, 4$)}
  \label{tab:wave-gen-params}
  \centering
  \begin{tabular}{lllll}
       \hline
       \textbf{Process}                         & $\mu$                        & $\sigma$  & $\theta$ & $R_{0}$
    \\ \hline
       $\{ R_{t}^{\operatorname{time}} \}_{t}$  & $1$                          & $2^{-8}$  & $2^{-8}$ & $0$
    \\ $\{ R_{t}^{\operatorname{amp} k} \}_{t}$ & $\mu^{\operatorname{amp} k}$ & $2^{-8}$  & $2^{-8}$ & $\mu^{\operatorname{amp} k}$
    \\ $\{ R_{t}^{\operatorname{ph} k} \}_{t}$  & $\mu^{\operatorname{ph} k}$  & $2^{-10}$ & $2^{-8}$ & $\mu^{\operatorname{ph} k}$
    \\ $\{ R_{t}^{\operatorname{noise}} \}_{t}$ & $0$                          & $2^{-6}$  & $2^{-8}$ & $0$
    \\ $\{ N_{t} \}_{t}$                        & $0$                          & $2^{-4}$  & N/A      & N/A
    \\ \hline
  \end{tabular}
\end{table}
The means of the processes for amplitude and phase variation are sampled according to the following law:
\[
  \log_{2} \mu^{\operatorname{amp} k} \sim U(-1, 1)
  \quad \text{and} \quad
  \mu^{\operatorname{ph} k} \sim U(0, 1)
\]
for $k = 1, 2, 3, 4$, where $U(a, b)$ denotes the uniform distribution on the interval $[a, b)$.
They remain constant throughout the wave and determine the overall shape of the wave.

\subsubsection{Generated anomalies}
Based on the parameters and processes employed by the wave generator, we inject the following four types of anomalies or noise:
\begin{description}
  \item[Amplitude anomalies]
    The amplitude process\linebreak $\{ R_{t}^{\operatorname{amp} k} \}_{t}$ of one of the frequency components (i.e., for a single $k \in \{ 1, 2, 3, 4 \}$) is increased by $a$, where $a$ is randomly sampled for each anomaly according to the law $\log_{2} a \sim U(1, 2)$.
  \item[Phase anomalies]
    The phase process $\{ R_{t}^{\operatorname{ph} k} \}_{t}$ of one of the frequency components is changed.
    The amount of change is randomly sampled for each anomaly from the distribution $U(1/4, 3/4)$ resulting in a random phase change of at least $90^{\circ}$ and at most $270^{\circ}$.
  \item[Pulse anomalies]
    A pulse of random amplitude is added onto the wave.
    For each anomaly, the amplitude $p$ of the pulse is randomly sampled according to the law $\log_{2} p \sim U(2, 4)$ and the pulse width is a random integer drawn from the interval $[2^{5}, 2^{6})$.
  \item[White noise]
    The white noise process $\{ N_{t} \}_{t}$ is amplified by a factor $\alpha$ which is randomly sampled for each anomaly according to the law $\log_{2} \alpha \sim U(2, 6)$.
\end{description}
For each wave, a segment of $2^{16}$ samples is generated.
Then $16$ segments, each consisting of $2^{12}$ samples are generated, the last $2^{11}$ samples of which the anomaly or noise is injected into.
For the evaluation, we use $24$ generated waves, resulting in a number of $290$ anomalies and $94$ waves with increased white noise in the test dataset.

\subsubsection{Input data and period detection}
The generated waves are considered in $24$ groups, where each group consists of a normal wave recorded over $2^8=256$ periods and further recordings, each injected with a single type of anomaly with a normal start-up time of at least $2^{11}=2048$ time units (i.e.~the first entrance time of anomalies following the respective normal wave is to the right of the time stamp $2^{11}=2048$).
In each group, we take the first $7/8$ and the remainder of the normal wave for training and validation, respectively, and subsequently test the trained classifier on the respective anomaly-injected test recordings.

Since the simulated waves contain interference in the time component which results in random period lengths $s$, we again make use of the period detector decribed in Appendix~\ref{pds} using the parameters specified in Table~\ref{tab:wave-pds-params}.
\begin{table}[htbp!]
  \caption{Parameters for period detector on wave dataset}
  \label{tab:wave-pds-params}
  \centering
  \begin{tabular}{lr}
       \hline
       \textbf{Parameter}                              & \textbf{Value}
    \\ \hline
       prefilter window half-length $n$                &   $8$
    \\ minimum base period length $s_{\min}$           & $240$
    \\ maximum base period length $s_{\max}$           & $272$
    \\ maximum period length deviation factor $\sigma$ & $1/4$
    \\ reference window half-length factor $\lambda$   & $1/3$
    \\ \hline
  \end{tabular}
\end{table}
Note that in contrast to the treatment of ECG data, in each data group the reference window is selected among the subpatterns extracted from the \emph{training data}.

By construction, the average period length equals $\overline{s}=2^8=256$ time untis.

\beginnew
\subsubsection{Hyperparameters}
Throughout the entire training, we set the maximum number of classes and allowed per-class margin of error to $n^*_0:=10$ and $\alpha :=2^{-6}$, respectively.

As presented in Table~\ref{tab:net-layout}, for each of the $24$ waves the corresponding classifier neural nets are all endowed with $M^{(0)}=1$ channel and $M^{(1)}=6$ channels on the first and second convolutional layers, respectively, where max pooling of size $R^{(1)}=3$ is applied between the covolutional layers, and two fully connected layers are set between the last convolutional layer and the output layer.
During the classifier selection process, the length $T^{(0)}$ of the input sequence, the sizes $S^{(0)}$, $S^{(2)}$ of the convolution kernels, and the size $N^{(3)}$ of the first fully connected layer vary proportionally to the sliding window length $T=\lfloor3\overline{s}/n_0\rfloor=\lfloor768/n_0\rfloor$ where $n_0$ runs over the values in $\{10(=n_0^*),8,6,4\}$ if not stopped earlier.
The size $N^{(4)}$ of the second fully connected layer is the geometric mean of the sizes $N^{(3)}$ and $N^{(5)}$ of its adjecent layers and the size $N^{(5)}$ of the output layer is equal to the current number of classes $n$ which runs over the values in $\{n_0,n_0-1,\ldots\}$ during the dynamic reclustering.

The ADAM optimiser with learning rate $\gamma=0.01$ is employed for training with SGD.
The mini-batch sizes are dynamically increased after every $2$ or $3$ epochs from $40$ to $360$.
\endnew

\beginnew
\section{Experimental results}\label{sec:results}

In this section we present the empirical results of the treatment of the example datasets given in Section~\ref{sec:example datasets} following our general phase classification scheme described in Section~\ref{sec:concept}.
Here we provide both the results of selecting and training the optimal classifier neural networks and the results of anomaly detection obtained by evaluating the trained classifier neural networks on the test data (recall Section~\ref{sec:anom-detect}).
\endnew

\subsection{Cardiology dataset}\label{sec:result-cardio}
\begin{figure}[htbp!]
  \centering
  \includegraphics[width=.85\linewidth]{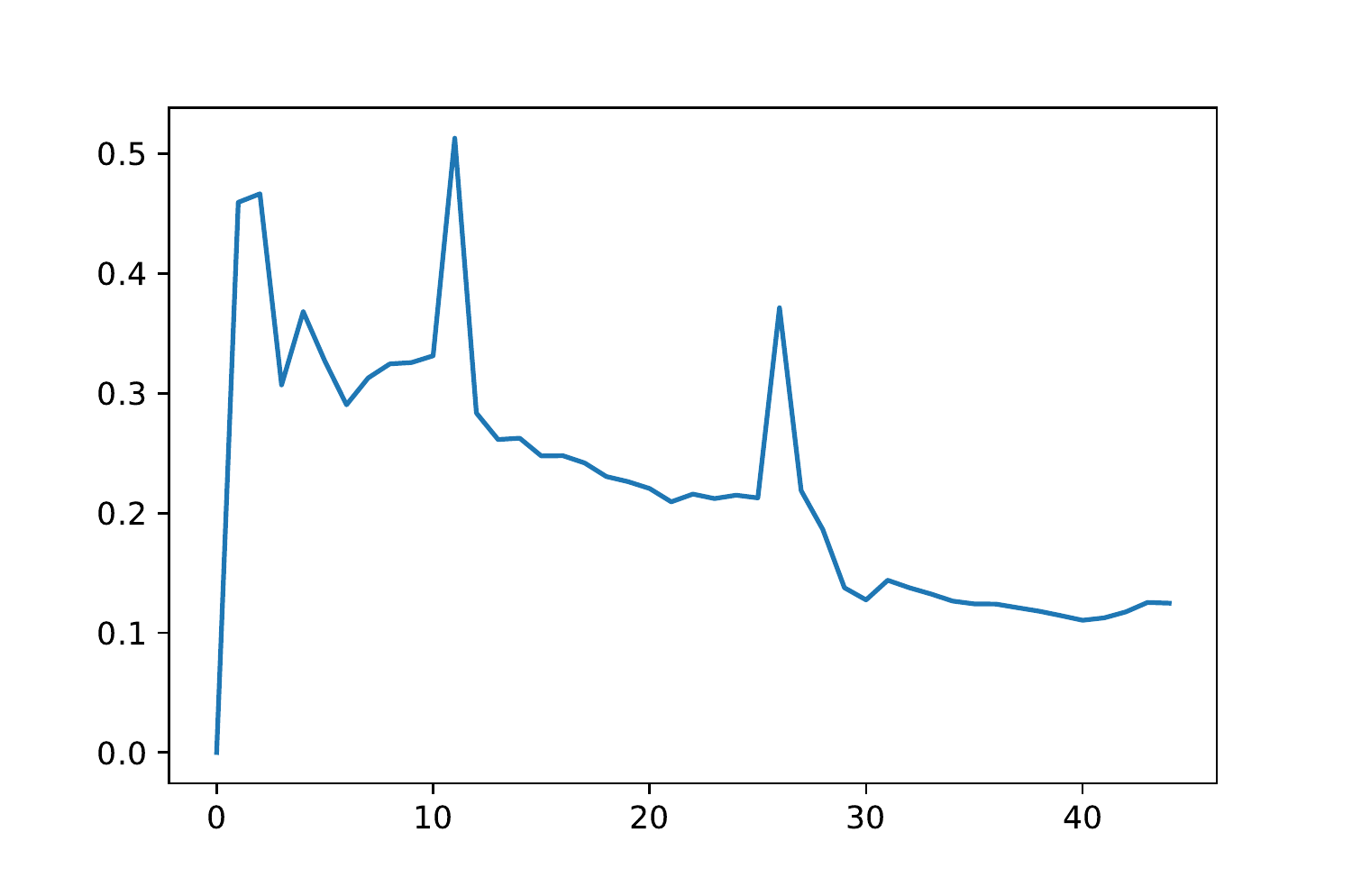}
  \caption{Validation loss over $44$ epochs for training the classifier neural nets with $n_0=6$, $n^{n_0}=4$ }
  \label{fig:Figure-loss-ECG}
\end{figure}

\beginnew
The ultimate classifier resulting from the dynamic model selection process turns out to be a classifier neural network corresponding to initial number of classes $n_0=6$ and final number of classes $n^{n_0}=4$, cf.~Table~\ref{tab:net-layout-cardio} for the layout of the final CNN.
\endnew
\begin{table}[htbp!]
  \caption{\beginnew Layers of final classifier neural network for ECG dataset\endnew}
  \label{tab:net-layout-cardio}
  \centering
  \begin{tabular}{lll}
       \hline
         & \textbf{Layer Type} & \textbf{Sizes}
    \\ \hline
       0 & convolutional   & $M^{(0)} = 15$, $T^{(0)} = 17$,
    \\   &                 & $S^{(0)} = 7$
    \\ 1 & max pooling     & $M^{(1)} = 90$, $T^{(1)} = 17$,
    \\   &                 & $R^{(1)} = 3$
    \\ 2 & convolutional   & $M^{(2)} = 90$, $T^{(2)} = 6$,
    \\   &                 & $S^{(2)} = 5$
    \\   &                 & $M^{(3)} = 270$, $T^{(3)} = 6$
    \\ 3 & fully connected & $N^{(3)} = 1620$
    \\ 4 & fully connected & $N^{(4)} = 80$
    \\ 5 & output          & $N^{(5)} = 4$
    \\ \hline
  \end{tabular}
\end{table}
The label history recorded along with the dynamic reclustering is shown in Table~\ref{tab:label history-ECG}.
\begin{table}[htbp!]
  \caption{Label History}
  \label{tab:label history-ECG}
  \centering
  \begin{tabular}{rrrr}
       \hline
       \multicolumn{1}{c}{\textbf{Epochs}} & \multicolumn{1}{c}{\textbf{Merge}} & \textbf{New Labels}
    \\ \hline
        $0$ --     $9$ & N/A        & $[0,1,2,3,4,5]$
    \\  $9$ $\to$ $10$ & $3$ to $2$ & $[0,1,2,2,4,3]$
    \\ $24$ $\to$ $25$ & $4$ to $2$ & $[0,1,2,2,2,3]$
    \\ $25$ --    $43$ & N/A        & $[0,1,2,2,2,3]$
    \\ \hline
  \end{tabular}
\end{table}
The average validation loss recorded during the training of the respective neural nets is presented in Figure~\ref{fig:Figure-loss-ECG}.
A training accuracy of $99\%$ and a validation accuracy of $96\%$ are achieved.

\begin{figure}[htbp!]
  \centering
  \begin{subfigure}{\linewidth}
    \includegraphics[width=.95\linewidth]{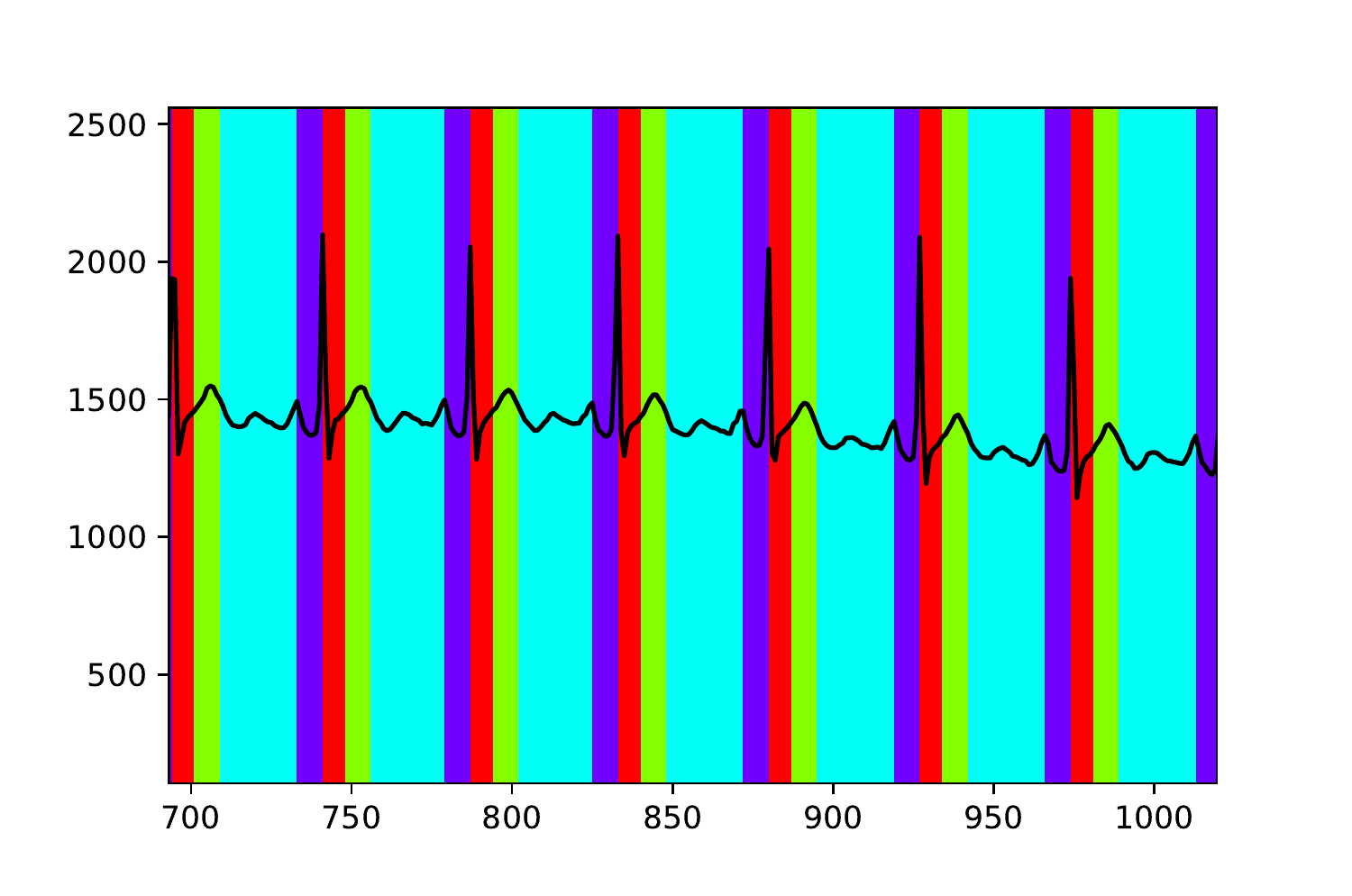}
  \end{subfigure}
  \begin{subfigure}{\linewidth}
    \includegraphics[width=.95\linewidth]{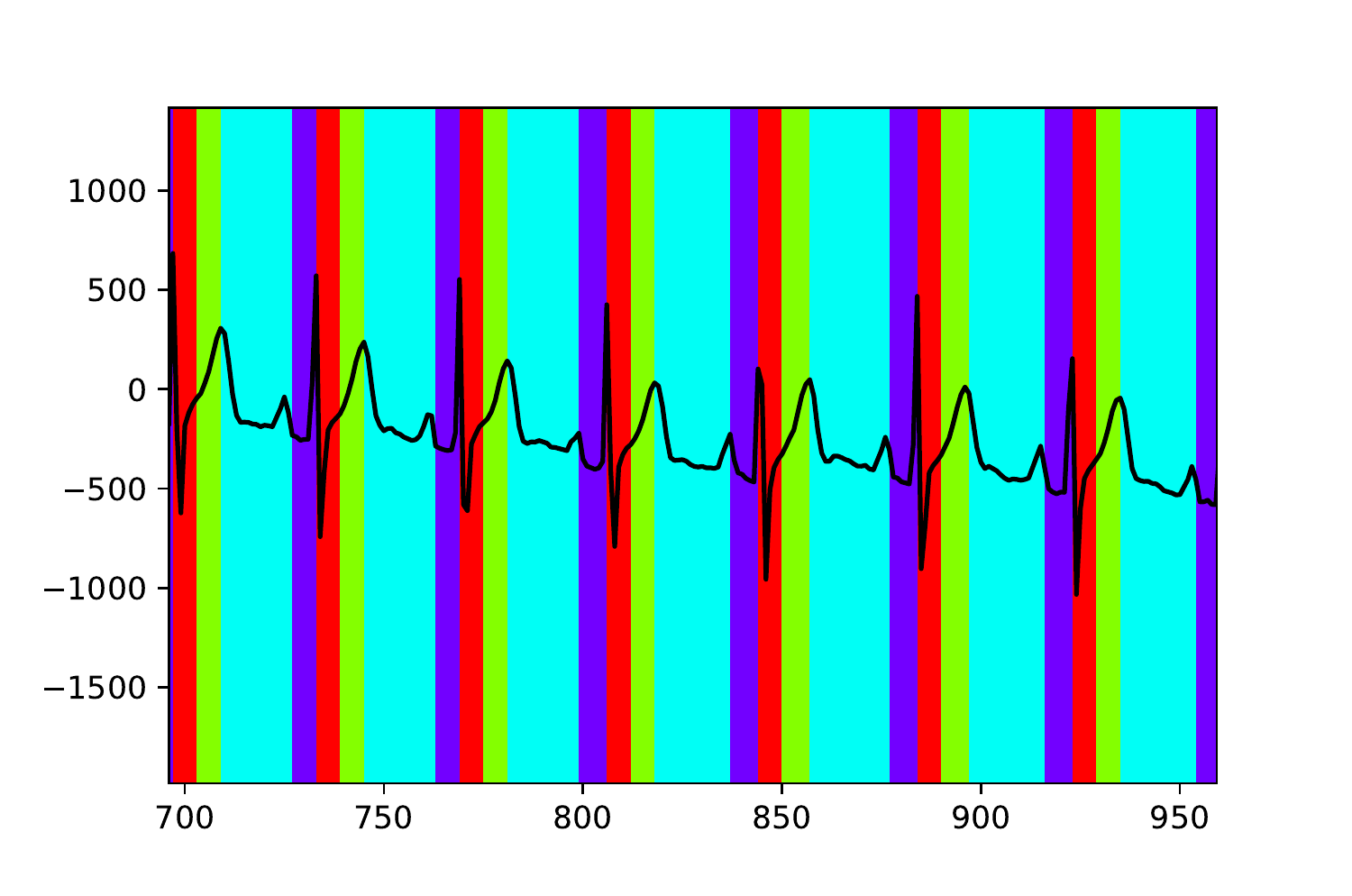}
  \end{subfigure}
  \begin{subfigure}{\linewidth}
    \includegraphics[width=.95\linewidth]{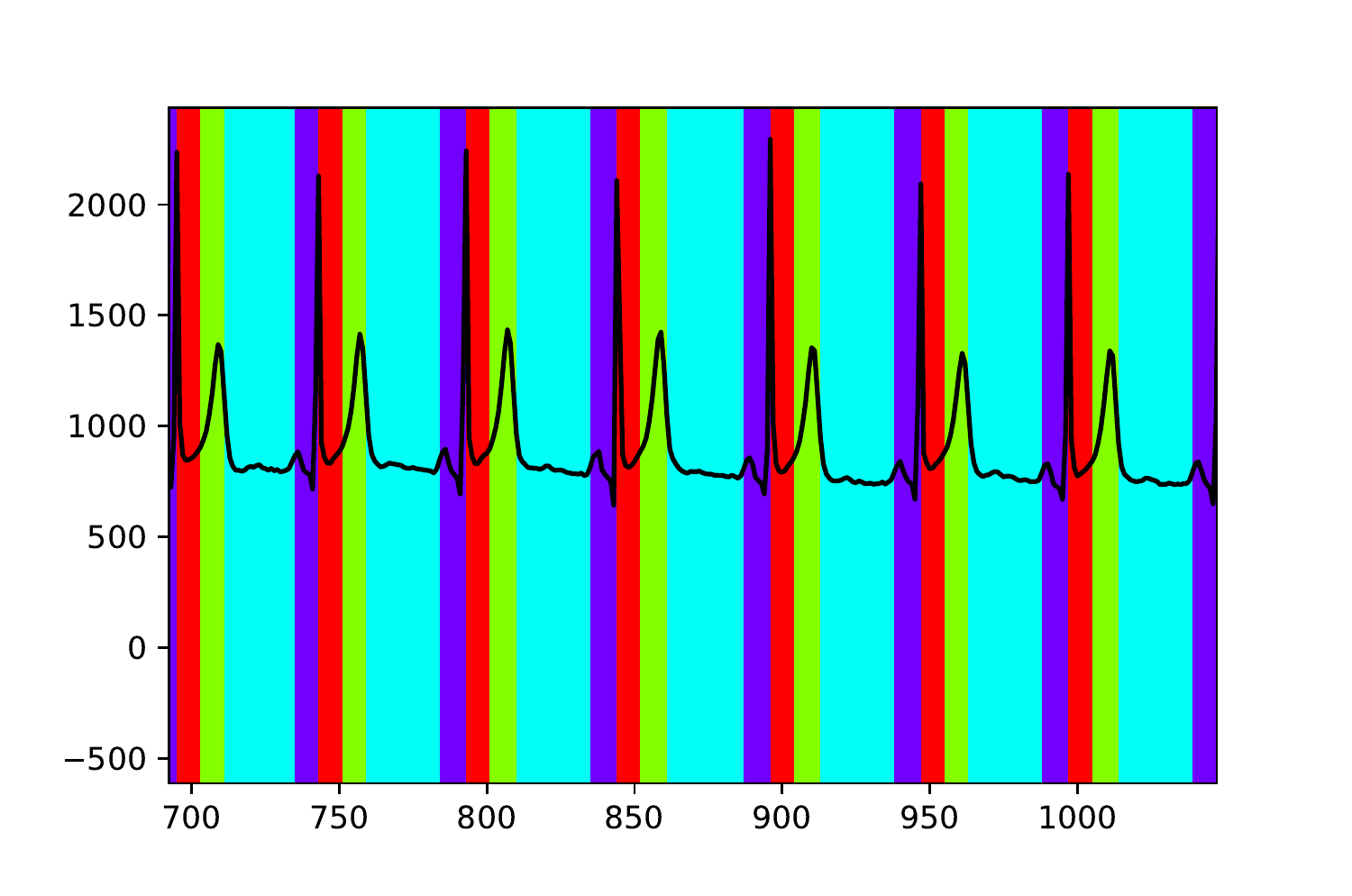}
  \end{subfigure}
  \caption{Classifier applied to patients from category `healthy control'}
  \label{fig:Figure-classes-ECG-healthy}
\end{figure}

\begin{figure}[htbp!]
  \centering
  \begin{subfigure}{\linewidth}
    \includegraphics[width=.95\linewidth]{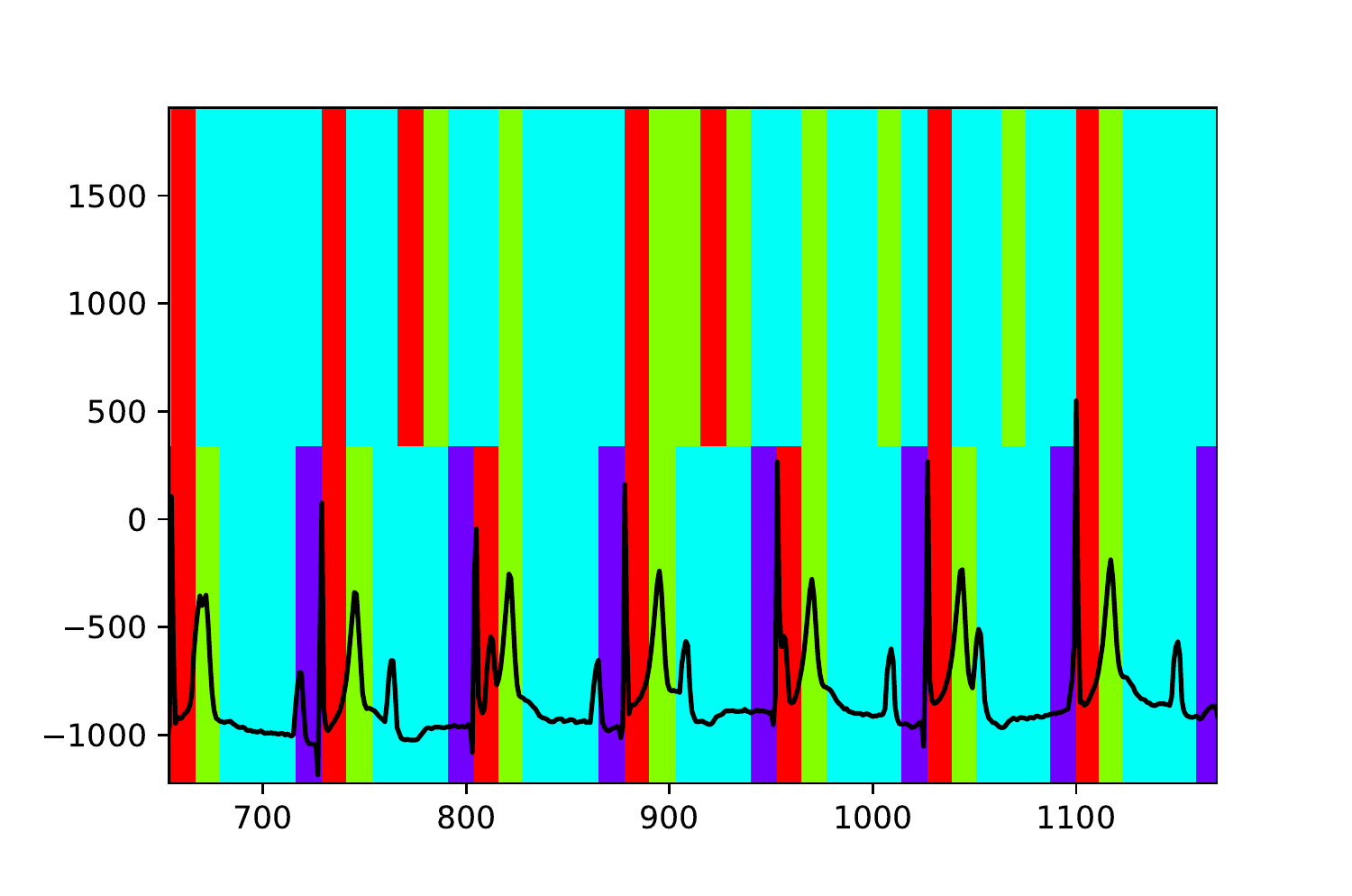}
    \caption{Classifier applied to a dysrthythmia patient}
  \end{subfigure}
  \begin{subfigure}{\linewidth}
    \includegraphics[width=.95\linewidth]{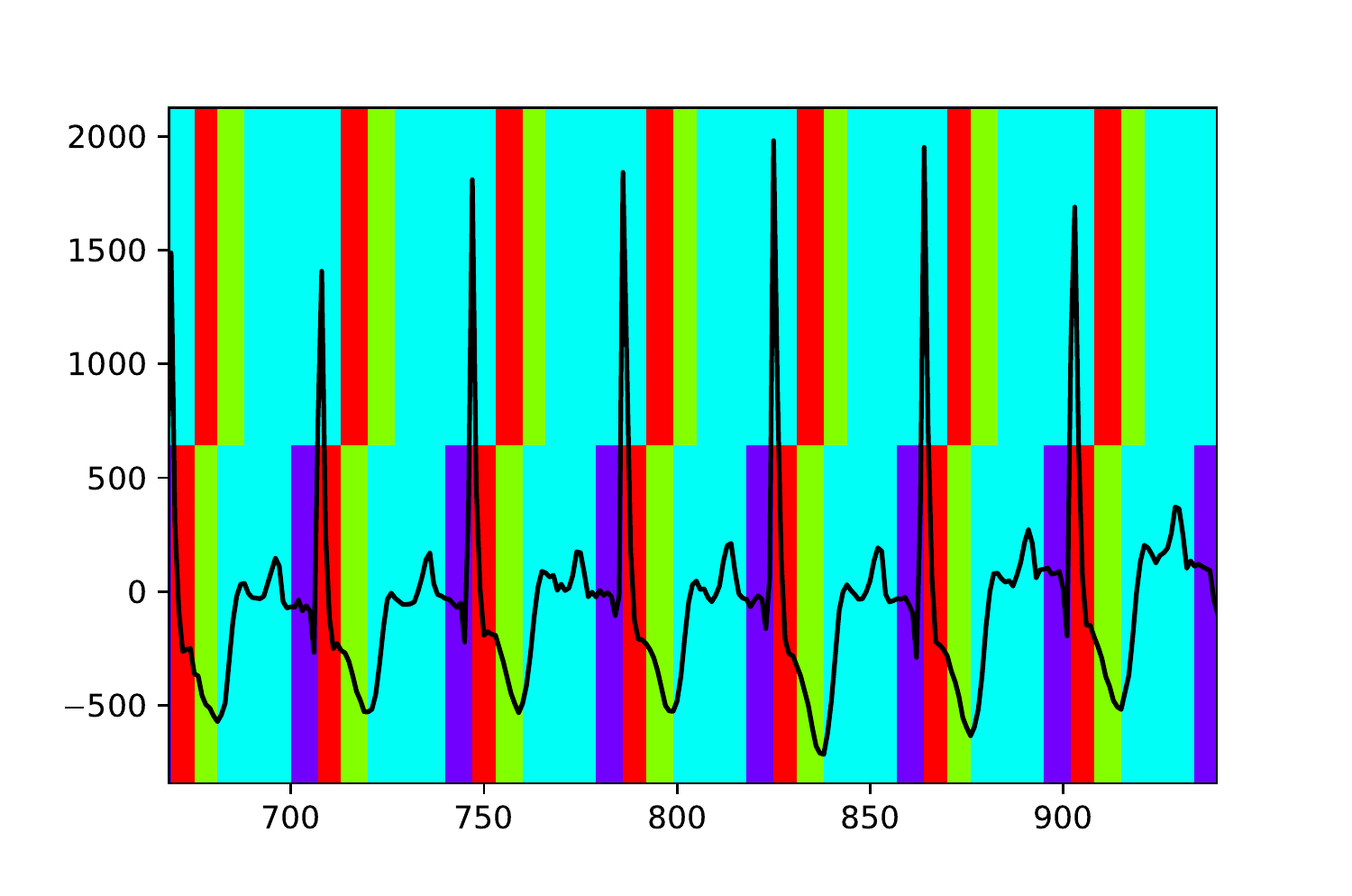}
    \caption{Classifier applied to a valvular-heart-disease patient}
  \end{subfigure}
  \begin{subfigure}{\linewidth}
    \includegraphics[width=.95\linewidth]{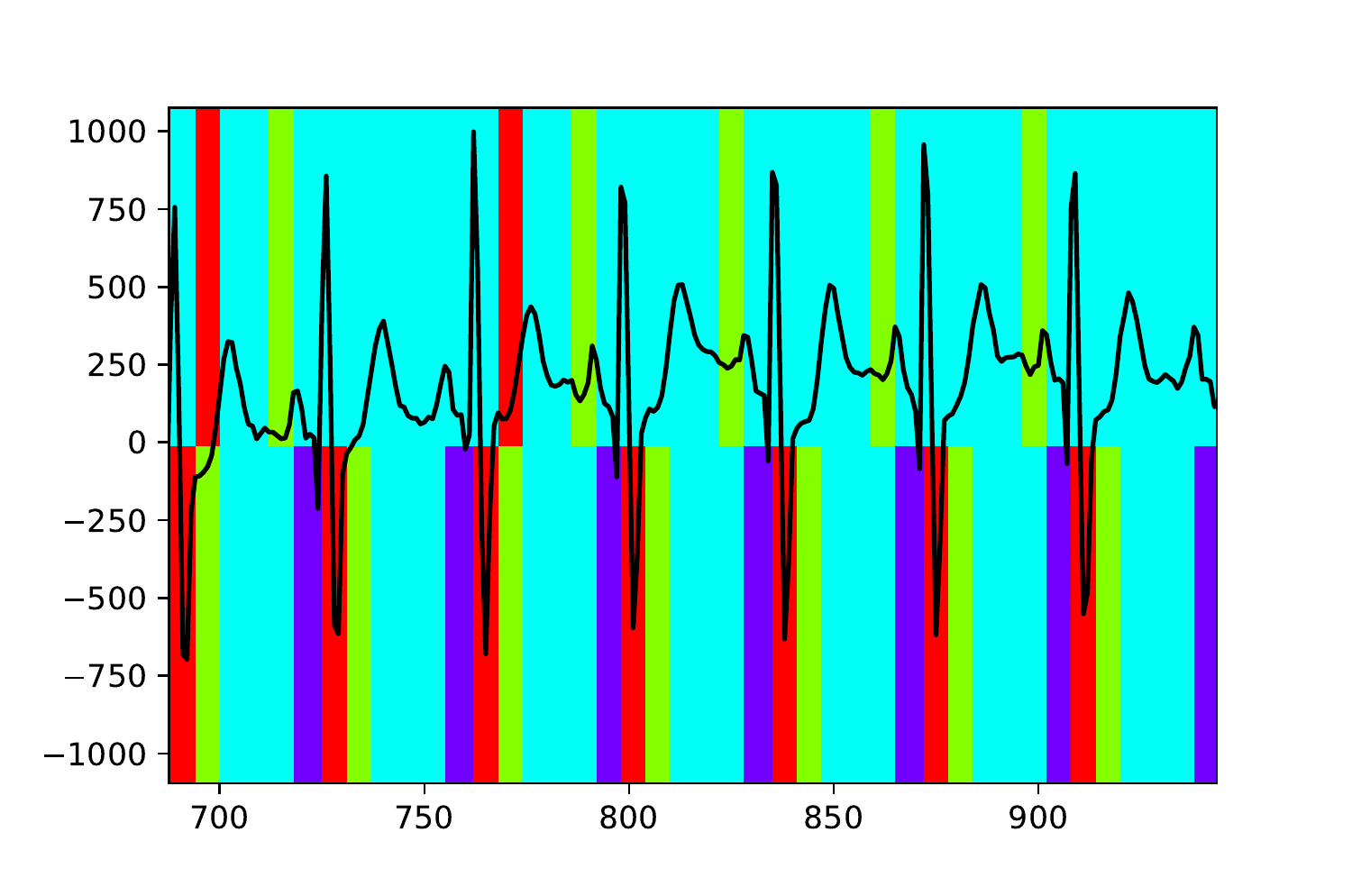}
    \caption{Classifier applied to a myocardial-infarction patient}
  \end{subfigure}
  \caption{Classifier applied to ill patients}
  \label{fig:Figure-classes-ECG-ill}
\end{figure}

Figures~\ref{fig:Figure-classes-ECG-healthy} and~\ref{fig:Figure-classes-ECG-ill} illustrate the result of testing the trained classifier on three patients from the category `healthy control' and three ill patients: the measurements on feature `i' from the test patients are presented in a temporal resolution of $20 \, \text{ms}$ and the bars in the upper and lower halves of the figures refer to the predicted classes and the true labels of the segments from the considered test signals, respectively.\footnote{Note that here and in the sequel the coloured bars in these diagrams are always plotted between the beginnings of adjacent segments to be classified, thus only covering approximately the first third of each segment.}

\begin{figure}[htbp!]
  \centering
  \includegraphics[width=.85\linewidth]{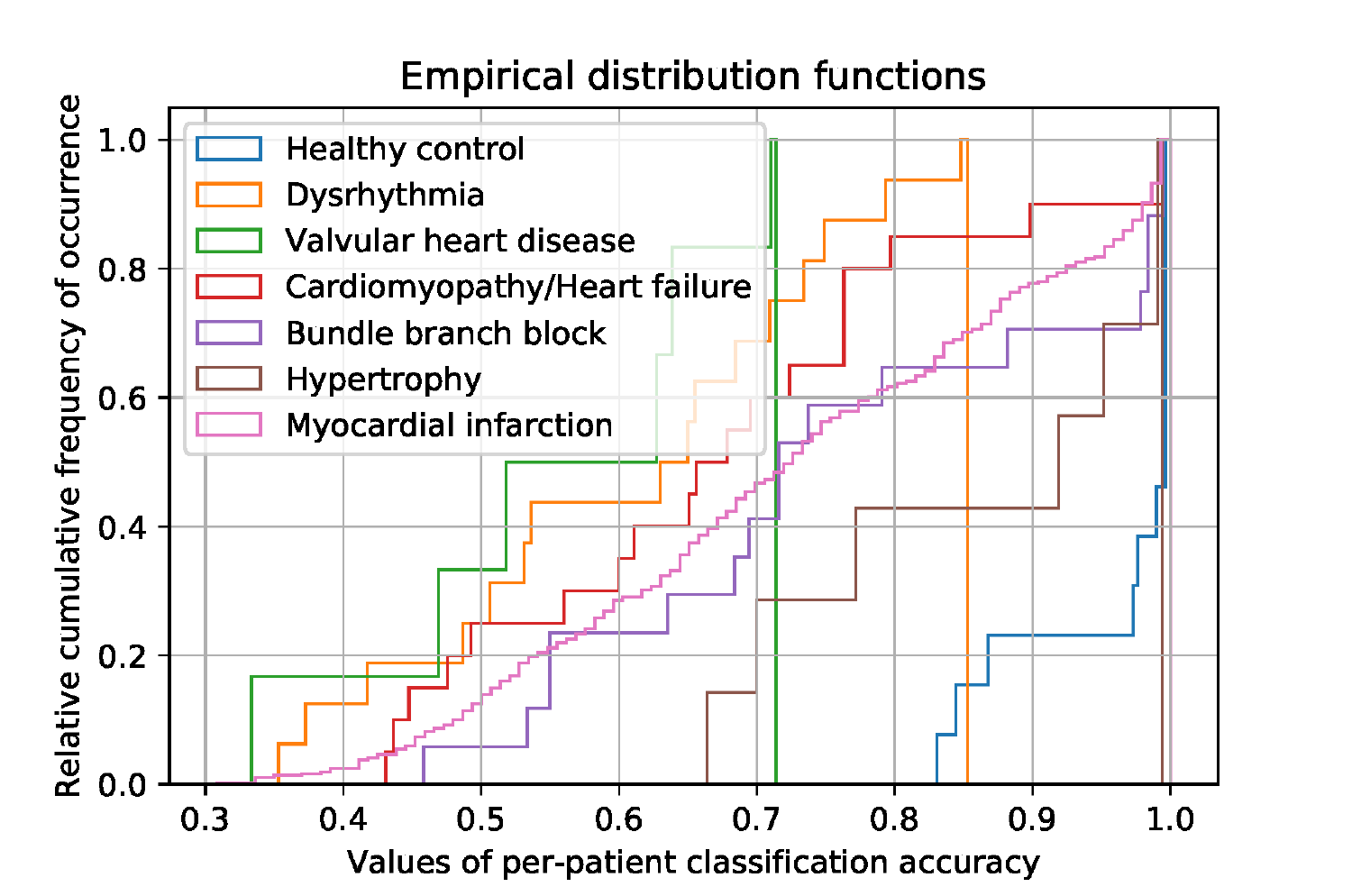}
  \caption{Distribution of per-patient classification accuracy evaluated on test patients from different categories}
  \label{fig:Figure-hist}
\end{figure}

Figure~\ref{fig:Figure-hist} presents a statistical evaluation of the per-patient test results on patients from the $7$ most recorded categories in the considered database: `dysrhythma', `valvular heart disease', `cardiomyopathy\slash heart failure', `bundle branch block', `hypertrophy', `myocardial infarction', and `healthy control'.
The lines in different colors represent the empirical distribution functions of the per-patient classification accuracy from the aforementioned categories.
Observe that the blue line related to healthy patients is located in the bottom right corner of the diagram, to the left of which all other lines corresponding to ill patients are centred (cf.~the median for each category), which enables us to distinguish ill patients from healthy patients in some cases.
For instance, according to the figure, if we take the average validation accuracy of $96\%$ as the threshold for the per-patient classification accuracy, all test patients from the categories `dysrhythma' and `valvular heart disease', $90\%$ and nearly $85\%$ of the patients from the categories `cardiomyopathy\slash heart failure' and `myocardial infarction', respectively, and over $70\%$ of the patients from the categotries `bundle branch block' and `hypertrophy' will be considered as anomalies, whereas up to three false positive results ($25\%$ of) all tested patients from the category `healthy control' will be assessed as normal.
Since the sample sizes provided for the individual categories vary a lot (e.g.~there are $148$ subjects for `myocardial infarction' whereas the entire category `healthy control' consists of only $52$ subjects including training, validation and test data applied in our context), we are not in the position to make a general statement on the choice of an ideal threshold value.
Table~\ref{tab:res-ECG} provides a statistical evaluation of the per-disease average classification accuracy.
\begin{table}[htbp!]
  \caption{Results of per-disease classification accuracy}
  \label{tab:res-ECG}
  \centering
  \begin{tabular}{rrrr}
       \hline
       \multicolumn{1}{c}{\textbf{Disease}} & \multicolumn{1}{c}{\textbf{Classification}}
    \\                                      & \multicolumn{1}{c}{\textbf{Accurracy}}
    \\ \hline
       Valvular heart disease        & $56\%$
    \\ Dysrhythmia                   & $60\%$
    \\ Cardiomyopathy/Heart failure  & $64\%$
    \\ Myocardial infarction         & $73\%$
    \\ Bundle branch block           & $76\%$
    \\ Hypertrophy                   & $86\%$
    \\ \hline
       Healthy control               & $97\%$
    \\ \hline
  \end{tabular}
\end{table}
It turns out that the category `healthy control' presents by far the best test result compared to all other categories related to heart disease (anomaly).

Note that our anomaly detection scheme does not incorporate any specific cardiological knowledge.
It gives an indication whether a patient may be ill or not, it detects deviations from the known healthy data and does not classify the diseases separately.
It also only gives a statistical indication, which is a result somewhat similar to the one reported in~\cite{DYM09} where it was observed that the ECGs of ill patients showed deviations in certain affine dependencies usually present between the 12-lead and 3-lead ECGs of healthy patients.

\subsection{SCADA dataset}
\begin{figure}[htbp!]
  \centering
  \includegraphics[width=.85\linewidth]{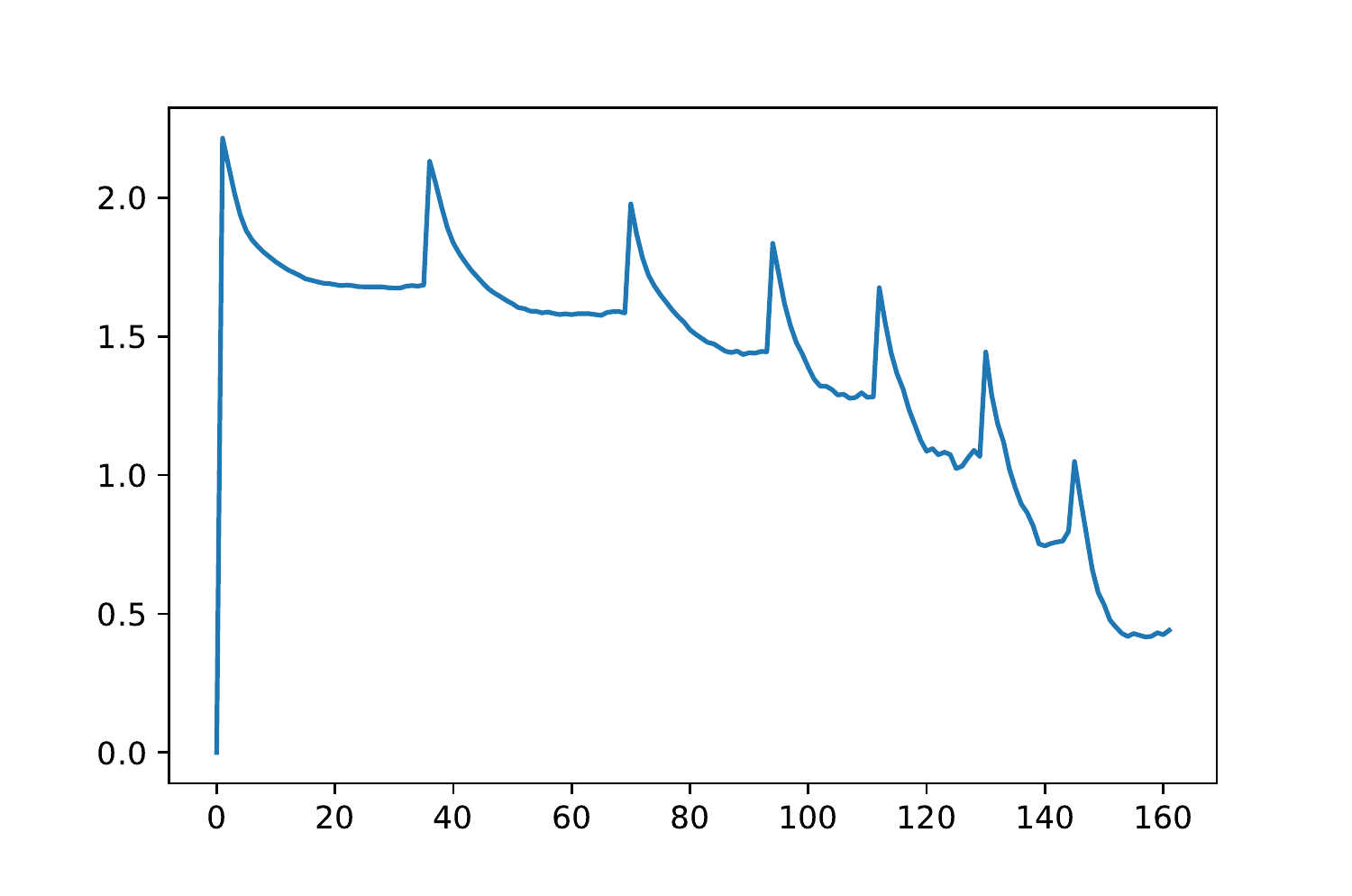}
  \caption{Validation loss over $161$ epochs for training the classifier neural nets with $n_0=10$, $n^{n_0}=4$}
  \label{fig:Figure-loss}
\end{figure}

\beginnew
The final classifier determined by means of the dynamic model selection scheme uses $n_0=10$ and $n^{n_0}=4$, cf.~Table~\ref{tab:net-layout-SCADA} for the layout of the final CNN.
\endnew
\begin{table}[htbp!]
  \caption{\beginnew Layers of final classifier neural network for SCADA dataset\endnew}
  \label{tab:net-layout-SCADA}
  \centering
  \begin{tabular}{lll}
       \hline
         & \textbf{Layer Type} & \textbf{Sizes}
    \\ \hline
       0 & convolutional   & $M^{(0)} = 4$, $T^{(0)} = 3$,
    \\   &                 & $S^{(0)} = 3$
    \\ 1 & max pooling     & $M^{(1)} = 24$, $T^{(1)} = 3$,
    \\   &                 & $R^{(1)} = 1$
    \\ 2 & convolutional   & $M^{(2)} = 24$, $T^{(2)} = 3$,
    \\   &                 & $S^{(2)} = 3$
    \\   &                 & $M^{(3)} = 72$, $T^{(3)} = 3$
    \\ 3 & fully connected & $N^{(3)} = 216$
    \\ 4 & fully connected & $N^{(4)} = 29$
    \\ 5 & output          & $N^{(5)} = 4$
    \\ \hline
  \end{tabular}
\end{table}
The respective label history recorded during the dynamic reclustering and the evolution of the average validation loss are presented in Table~\ref{tab:label history} and Figure~\ref{fig:Figure-loss}, respectively.
\begin{table}[htbp!]
  \caption{Label History}
  \label{tab:label history}
  \centering
  \begin{tabular}{rrrr}
       \hline
       \multicolumn{1}{c}{\textbf{Epochs}} & \multicolumn{1}{c}{\textbf{Merge}} & \textbf{New Labels}
    \\ \hline
         $0$ --     $34$ & N/A        & $[0,1,2,3,4,5,6,7,8,9]$
    \\  $34$ $\to$  $35$ & $1$ to $4$ & $[0,4,2,3,4,5,6,7,8,1]$
    \\  $68$ $\to$  $69$ & $3$ to $4$ & $[0,4,2,4,4,5,6,7,3,1]$
    \\  $92$ $\to$  $93$ & $7$ to $4$ & $[0,4,2,4,4,5,6,4,3,1]$
    \\ $110$ $\to$ $111$ & $5$ to $4$ & $[0,4,2,4,4,4,5,4,3,1]$
    \\ $128$ $\to$ $129$ & $2$ to $4$ & $[0,4,4,4,4,4,2,4,3,1]$
    \\ $143$ $\to$ $144$ & $2$ to $4$ & $[0,2,2,2,2,2,2,2,3,1]$
    \\ $144$ --    $160$ & N/A        & $[0,2,2,2,2,2,2,2,3,1]$
    \\ \hline
  \end{tabular}
\end{table}

\begin{figure}[htbp!]
  \centering
  \includegraphics[width=.95\linewidth]{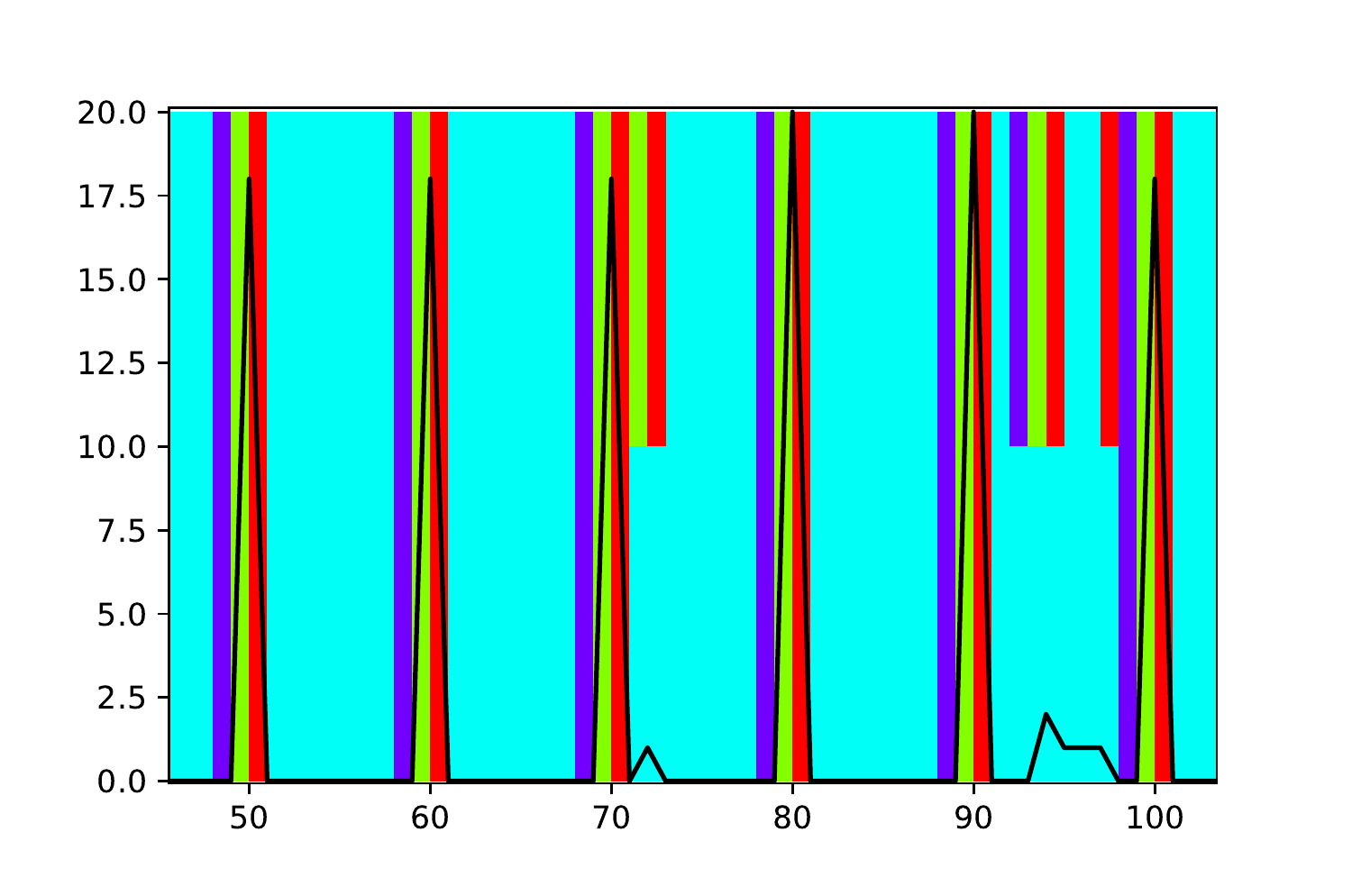}
  \caption{Classifier applied to test data}
  \label{fig:Figure-classes}
\end{figure}

In Figure~\ref{fig:Figure-classes}, the number of active port pairs extracted from `Test Data 1' is plotted against time (in seconds) and the bars in the upper and lower halves represent the classes predicted by our trained neural net and the true labels of the test segments, respectively; segments which result in prediction errors are considered anomalies.

The final results of our anomaly detection algorithm on the entire test data are summarised in Table~\ref{tab:res-modbus}.
\begin{table}[htbp!]
  \caption{Results of Intrusion Detection}
  \label{tab:res-modbus}
  \centering
  \begin{tabular}{rrrr}
       \hline
    \multicolumn{1}{c}{\textbf{Dataset}} & \textbf{Detection Rate} & \textbf{False Positives}
    \\ \hline
       Test Data 1 & $4/4$ & $0$
    \\ Test Data 2 & $3/3$ & $1\%$
    \\ Test Data 3 & $0/1$ & $8\%$
    \\ \hline
  \end{tabular}
\end{table}
In the first two (cleaner) test datasets with no or only a small amount of manual operations (noise), all cyber attacks in the test data are detected along with a single false positive detection (corresponding to $1\%$ false detection rate in `Test Data 1'), whereas the classifier tested on the last test dataset including a large amount of noise performs not as good, which is not surprising taking into account that only malicious activities but no manual operations or any other types of interference are labelled as anomalies and our time series analysis does not include the respective context consideration.

Indeed, the SCADA datasets which are applicable in our setting are quite small.
Due to the non-compatability between datasets with small and large amounts of noise (i.e.~non-intrusion anomalies appearing in the form of pulses), it is difficult to choose one suitable dataset for training and to test the intrusion detector on datasets with incompatible characteristics, e.g., it would be unfeasible to train an anomaly detector on one of the cleaner datasets and then test it against a noisy dataset, or vice versa.
For a more extensive treatment of anomaly detection of Type~B described in Section~\ref{sec:task} using a richer dataset and the corresponding results, cf.~Section~\ref{sec: waves} and Section~\ref{sec:result-waves}.

\subsection{Wave dataset}\label{sec:result-waves}
Overall, an average classification accuracy of $99\%$ is achieved on both training and validation data.

\begin{figure}[htbp!]
  \centering
  \includegraphics[width=.85\linewidth]{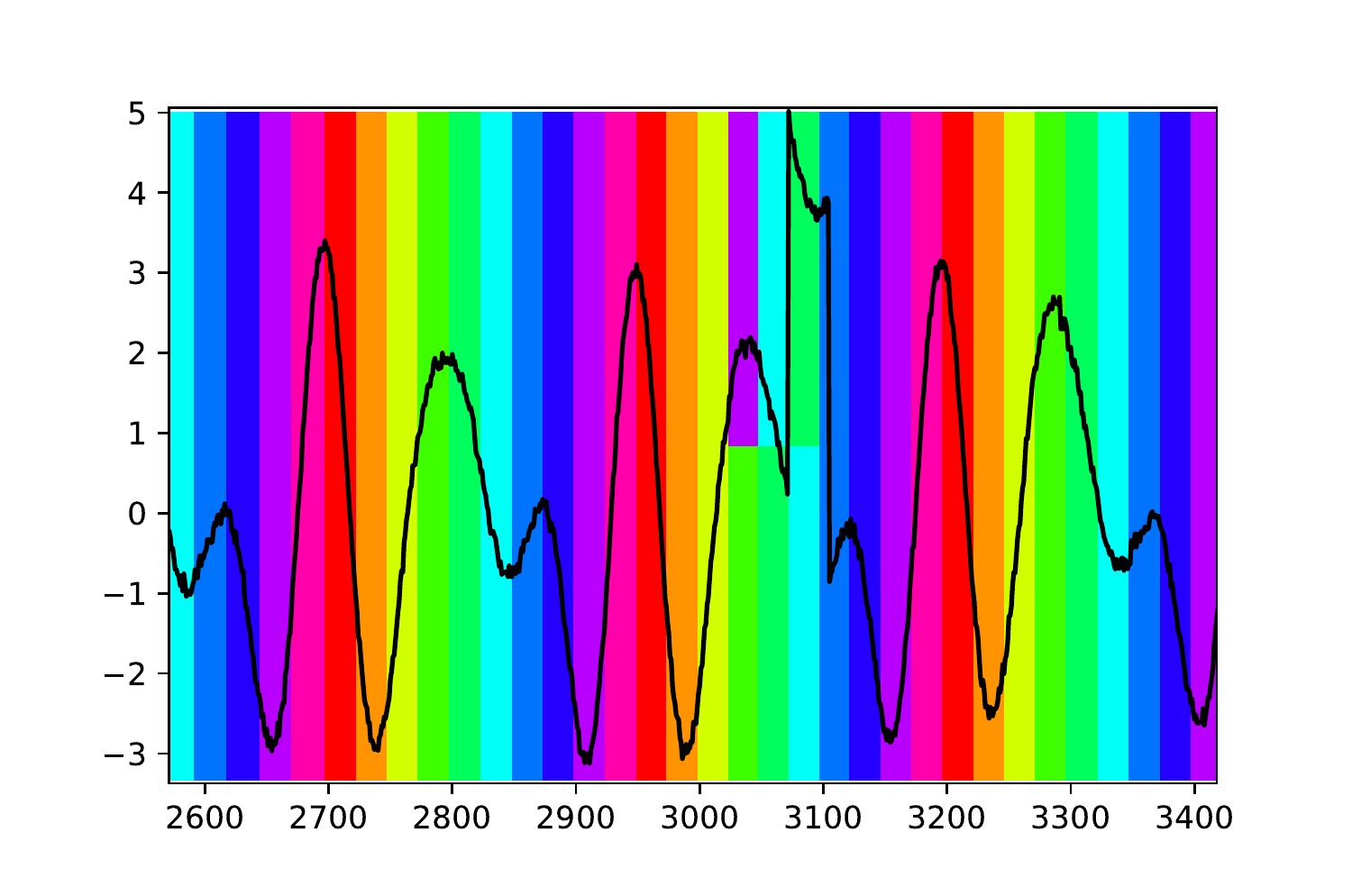}
  \caption{Classifier with $n_0=10$, $n^{n_0}=10$ applied to wave with pulse anomaly injected at time stamp 3072}
  \label{fig:Figure-pulse}
\end{figure}

\begin{figure}[htbp!]
  \centering
  \includegraphics[width=.85\linewidth]{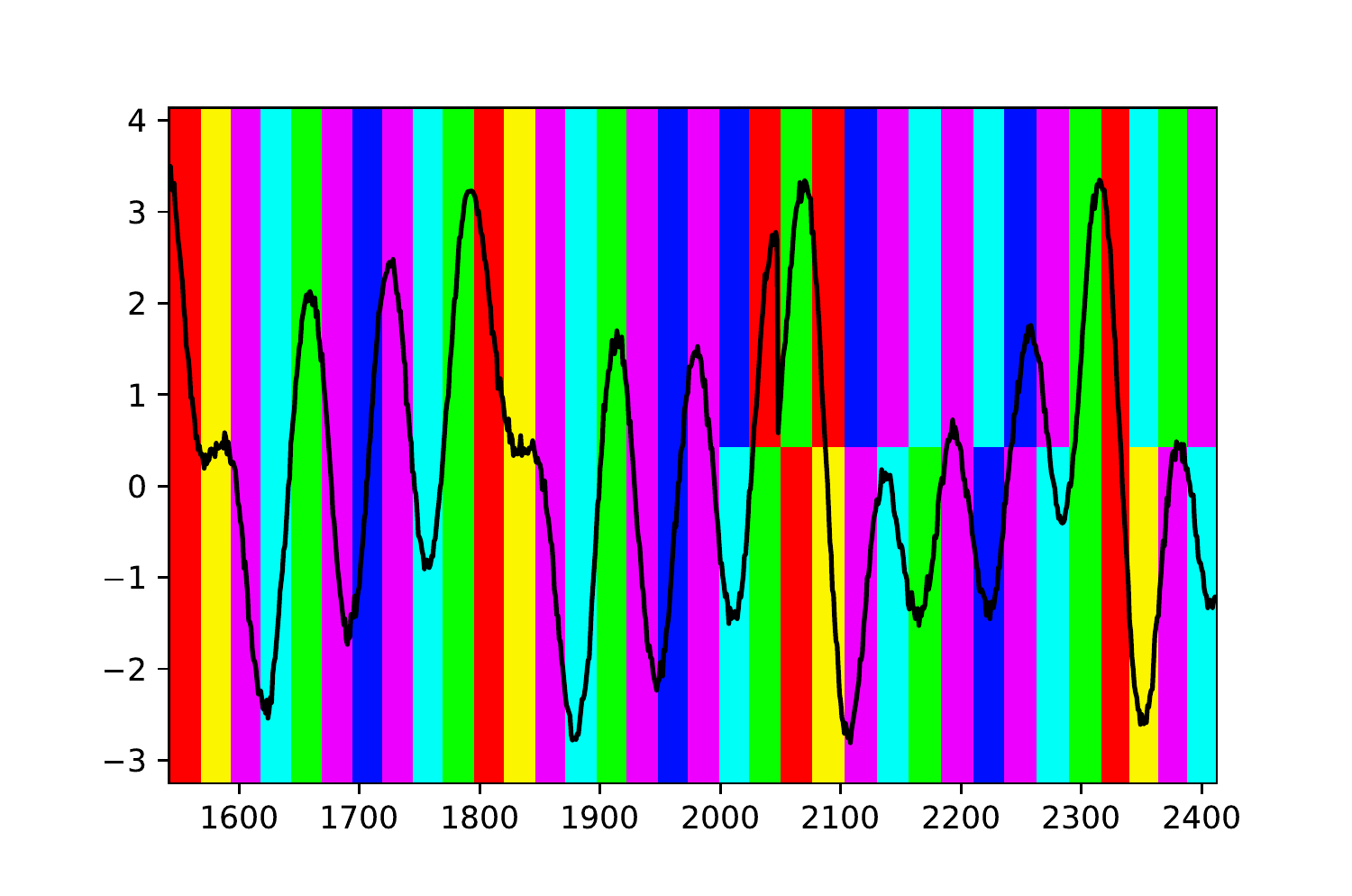}
  \caption{Classifier with $n_0=10$, $n^{n_0}=6$ applied to wave with abnormal phases starting at time stamp 2048}
  \label{fig:Figure-phases}
\end{figure}

\begin{figure}[htbp!]
  \centering
  \includegraphics[width=.85\linewidth]{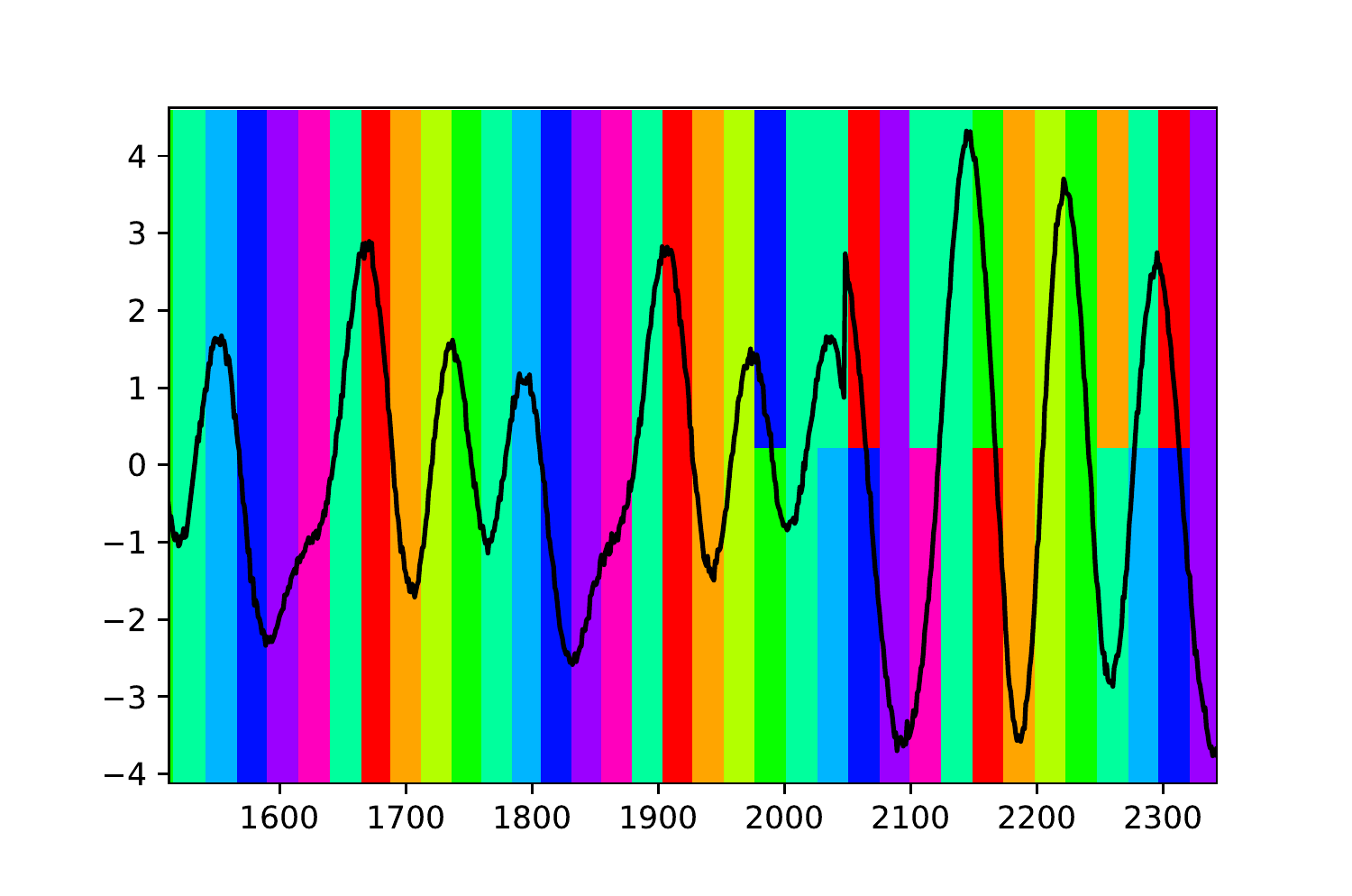}
  \caption{Classifier with $n_0=10$, $n^{n_0}=9$ applied to wave with abnormal amplitudes starting at time stamp 2048}
  \label{fig:Figure-amplitudes}
\end{figure}

\begin{figure}[htbp!]
  \centering
  \includegraphics[width=.85\linewidth]{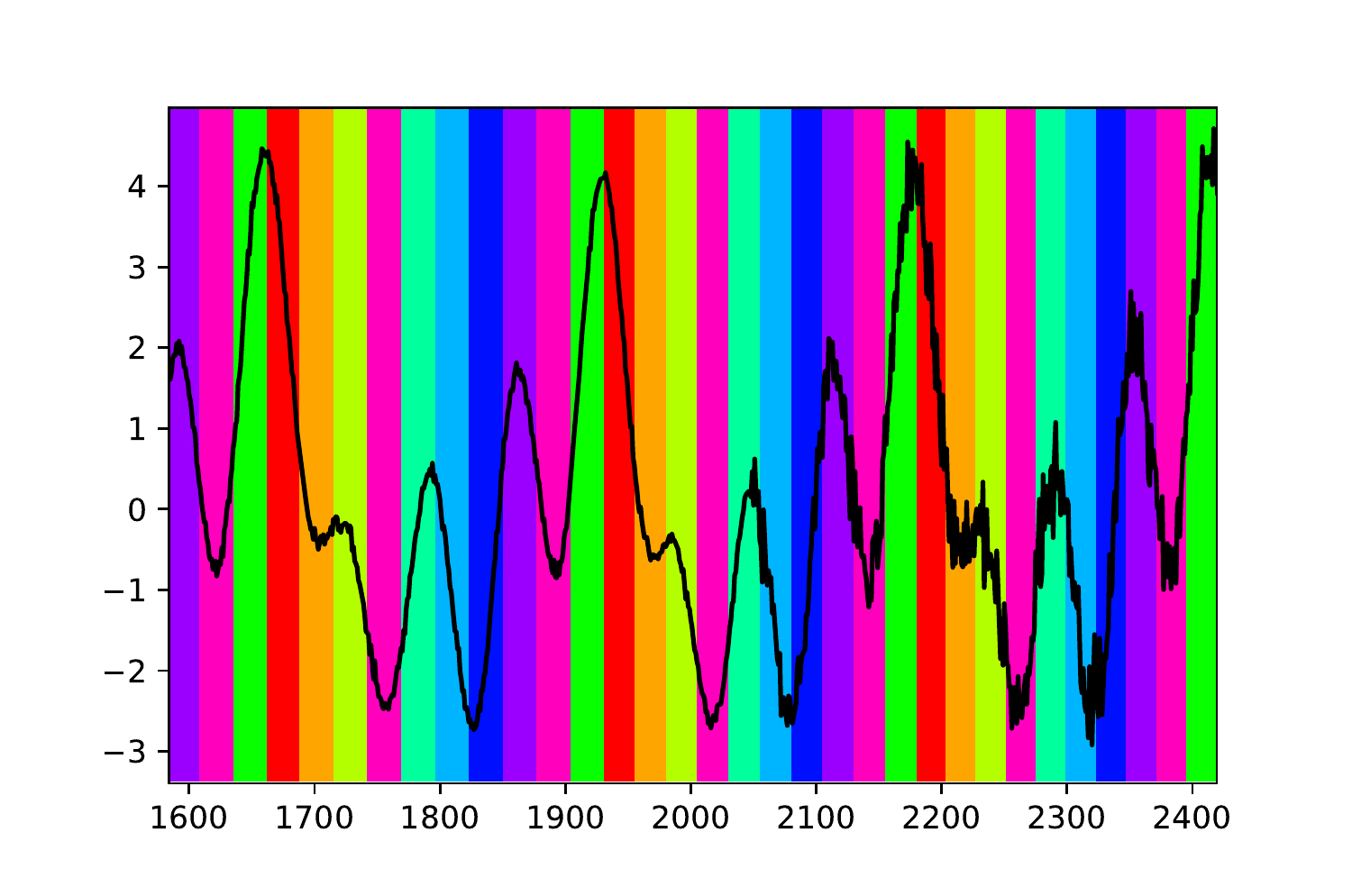}
  \caption{Classifier with $n_0=10$, $n^{n_0}=8$ applied to wave with slightly increased white noise ($\sigma=4.77$) starting at time stamp 2048}
  \label{fig:Figure-whitenoise-small}
\end{figure}

Figures~\ref{fig:Figure-pulse},~\ref{fig:Figure-phases},~\ref{fig:Figure-amplitudes} and Figure~\ref{fig:Figure-whitenoise-small} present the detection results of our classifiers trained by individual example waves and tested on segments injected with different types of anomalies and white noise, respectively.
Again, in each diagram the bars in the upper and lower halves refer to the predicted classes and true labels of the data from the test segments fed into the trained classifier, respectively.
Notice that in Figure~\ref{fig:Figure-whitenoise-small}, slightly increased white noise does not lead to any classification errors, which suggests some robustness property of our classifier against noise.

The final results of our anomaly detection algorithm tested on the $24$ groups of synthetic waves are shown in Table~\ref{tab:res-waves}.
\begin{table}[htbp!]
  \caption{Results of Anomaly Detection}
  \label{tab:res-waves}
  \centering
  \begin{tabular}{rrrr}
    \hline
       \multicolumn{1}{c}{\textbf{Type}} & \textbf{Detection Rate} & \%
    \\ \hline
       Phases                       & $83/85$     & $98\%$
    \\ Amplitudes                   & $102/102$   & $100\%$
    \\ Pulse                        & $102/103$   & $99\%$
    \\ \cline{2-3}
       Total Anomalies              & $287/290$   & $99\%$
    \\ \hline
       False Positives              & $115/26798$ & $0.43\%$
    \\ \hline
       White noise ($\sigma\leq 6$) & $11/19$     & $58\%$
    \\ White noise ($\sigma>6$)     & $64/75$     & $85\%$
    \\ \hline
  \end{tabular}
\end{table}
The amount of anomalies and white noise are obtained by counting the number of test waves injected with the respective type of interference, whereas the denominator for evaluating false positives equals the number of available prediction windows (test segments) in the clean test data.
Overall, our algorithm yields high detection rates of all types of injected anomalies ($99\%$ on average); the small rate of false positives ($<1\%$) confirms the model adequacy of our phase classification scheme; the low error rate in the presence of increased white noise shows the robustness of our classifier neural networks against noise to a certain extent.

\section{Conclusion}

In this paper, we proposed a novel approach to detecting anomalies in time series exhibiting periodic characteristics, where we applied deep convolutional neural networks for phase classification and automated phase similarity tagging.
We evaluated our approach on three example datasets corresponding to the domains of cardiology, industry, and signal processing, confirming that our method is feasible in a number of contexts.

\appendix
\section{Period detection scheme}\label{pds}

In this section, we provide the details for the period detection scheme used for the ECG and synthetic wave datasets.
This period detection scheme is primed using a reference signal $\{ Y^{\operatorname{raw}}_{t} \}_{t}$ and then applied to the actual input signal $\{ X^{\operatorname{raw}}_{t} \}_{t}$.
It is assumed that the input signals do not have a trend component, which can be achieved by a suitable transformation of the input signals, such as taking the first difference as in the ECG data case, cf.~Section~\ref{sec:cardio}.
The detection is now performed in the following steps:
\begin{enumerate}
  \item \label{pds-smooth}Smooth the signals by applying a rolling mean
  \item \label{pds-autocorr}Infer approximate base period using the autocorrelation of the reference signal
  \item \label{pds-signalpd}Detect peaks in the reference signal spaced approximately one base period apart using a simple peak detection logic
  \item \label{pds-refsegment}Take the average of segments around the detected peaks and find one reference segment which most closely matches this average
  \item \label{pds-xcorr}Cross-correlate the input signal with the reference segment
  \item \label{pds-xcorrpd}Detect peaks in the cross-correlation spaced approximately one base period apart using again the simple peak detection logic
\end{enumerate}

The steps are described in more detail in the following paragraphs.

Step~\ref{pds-smooth}:
The raw signals $\{ X^{\operatorname{raw}}_{t} \}_{t}$ and $\{ Y^{\operatorname{raw}}_{t} \}_{t}$ are subjected to a rolling mean filter, resulting in smoothed signals $\{ X_{t} \}_{t}$ and $\{ Y_{t} \}_{t}$, respectively, i.e.,
\[
     X_{t}
  := \frac{1}{2 n + 1} \sum_{k = -n}^{n} X^{\operatorname{raw}}_{t + k}
  , \enskip
     Y_{t}
  := \frac{1}{2 n + 1} \sum_{k = -n}^{n} Y^{\operatorname{raw}}_{t + k}
  .
\]
The window length $2 n + 1$ of this filter is chosen to provide just enough filtering to dampen some of the noise contained within the input signal.

\begin{figure}[htbp!]
  \centering
  \includegraphics[width=.85\linewidth]{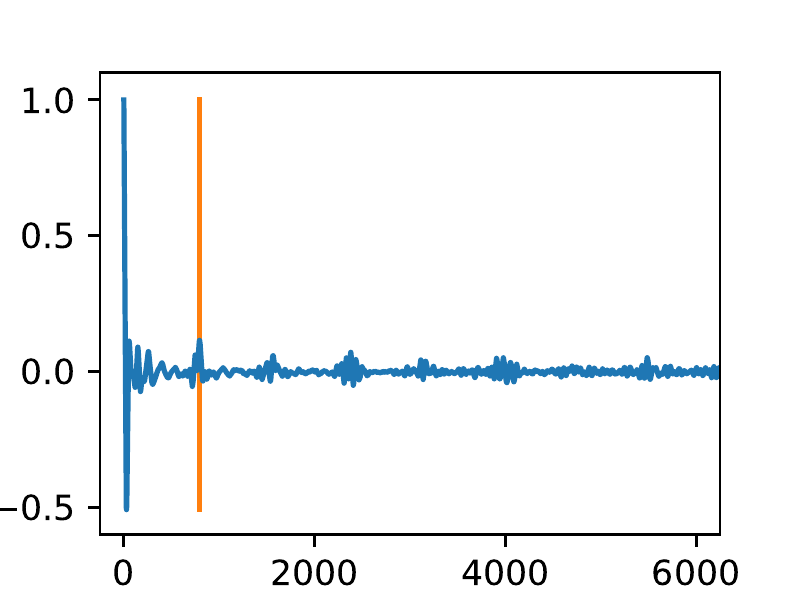}
  \caption{Autocorrelation function of one of the ECG database records}
  \label{fig:pds-autocorr}
\end{figure}

Step~\ref{pds-autocorr}:
The sample autocorrelation $\hat{\rho}_{\tau}^{Y}$ of the (smoothed) reference signal $\{ Y_{t} \}_{t}$ at lag $\tau$ is computed via
\[
     \hat{\rho}_{\tau}^{Y}
  := \frac{\hat{r}_{\tau}^{Y}}{\hat{r}_{0}^{Y}}
  \quad
  \text{for $\tau = 0, \ldots, N_{Y} - 1$}
\]
with
\[
     \hat{r}_{\tau}^{Y}
  := \frac{1}{N_{Y}} \sum_{t = 0}^{N_{Y} - 1 - \tau} (Y_{t + \tau} - \bar{Y}) (Y_{t} - \bar{Y})
\]
(cf.~\cite[2.1.5]{BJR08}), where $N_{Y}$ and $\bar{Y}$ denote the sample size and sample mean of the reference signal $Y$, respectively.
Now the mean period length $\hat{s}$ is inferred by taking the $\argmax$ of $\hat{\rho}_{\tau}$ restricted to some interval $[s_{\min}, s_{\max}]$, i.e.,
\[
     \hat{s}
  := \argmax_{s_{\min} \le \tau \le s_{\max}} \hat{\rho}_{\tau}^{Y}
  .
\]
A plot of an example autocorellation function is shown in Figure~\ref{fig:pds-autocorr}, the inferred mean period length is displayed by the vertical line.

Step~\ref{pds-signalpd}:
The reference signal is now fed into a simple peak detector which proceeds to inductively find peaks $T_{k}$ spaced approximately one base period apart via
\[
     T_{0}
  := \argmax_{0 \le t \le \lceil \hat{s} (1 + \sigma) \rceil} Y_{t}
  , \enskip
     T_{k + 1}
  := \argmax_{\hspace{-1.5em} T_{k} + \lfloor \hat{s} (1 - \sigma) \rfloor \le t \le T_{k} + \lceil \hat{s} (1 + \sigma) \rceil \hspace{-1.5em}} Y_{t}
  ,
\]
where $\sigma \in [0, 1)$ is a tolerance value to account for the variability of period lengths in the signals.

Step~\ref{pds-refsegment}:
The detector now extracts subpatterns $\{ U_{t}^{(k)} \}_{t = \lfloor -\hat{s} \lambda \rfloor}^{\lceil \hat{s} \lambda \rceil}$ from the reference signal $Y_{t}$ centred at the peaks $T_{k}$, i.e., $U_{t}^{(k)} = Y_{T_{k} + t}$.
Here, $\lambda \in (0, 1/2]$ is another tolerance parameter to mitigate the effects of period length variability.
Then the seasonal means
\[
  \bar{U}_{t} := \frac{1}{M} \sum_{k = 0}^{M - 1} U_{t}^{(k)}
\]
are computed.
Here $M$ denotes the total number of subpatterns.
Let now
\[
     k_{0}
  := \argmax_{k} \sum_{t = \lfloor -\hat{s} \lambda \rfloor}^{\lceil \hat{s} \lambda \rceil} U_{t}^{(k)} \bar{U}_{t}
  \quad \text{and} \quad
     U_{t}^{\operatorname{ref}}
  := U_{t}^{(k_{0})}
  .
\]
The choice of $k_{0}$ ensures that $\{ U_{t}^{\operatorname{ref}} \}_{t = \lfloor -\hat{s} \lambda \rfloor}^{\lceil \hat{s} \lambda \rceil}$ is the subpattern with maximum similarity to the mean $\{ \bar{U}_{t} \}_{t = \lfloor -\hat{s} \lambda \rfloor}^{\lceil \hat{s} \lambda \rceil}$ and is thus suited as a reference pattern.

Step~\ref{pds-xcorr}:
The reference pattern is now used for detecting the periods in the input signal by computing the cross-correlation function:
\[
     C_{\tau}
  := (X \star U^{\operatorname{ref}})_{\tau}
  =  \sum_{t = \lfloor -\hat{s} \lambda \rfloor}^{\lceil \hat{s} \lambda \rceil} X_{\tau + t} U^{\operatorname{ref}}_{t}
  , \enskip
  \tau \ge 0
\]

Step~\ref{pds-xcorrpd}:
Finally the simple peak detector from Step~\ref{pds-signalpd} is applied to the cross-correlation $\{ C_{\tau} \}_{\tau}$ to obtain the final segment beginnings.

\begin{figure}[htbp!]
  \centering
  \includegraphics[width=.85\linewidth]{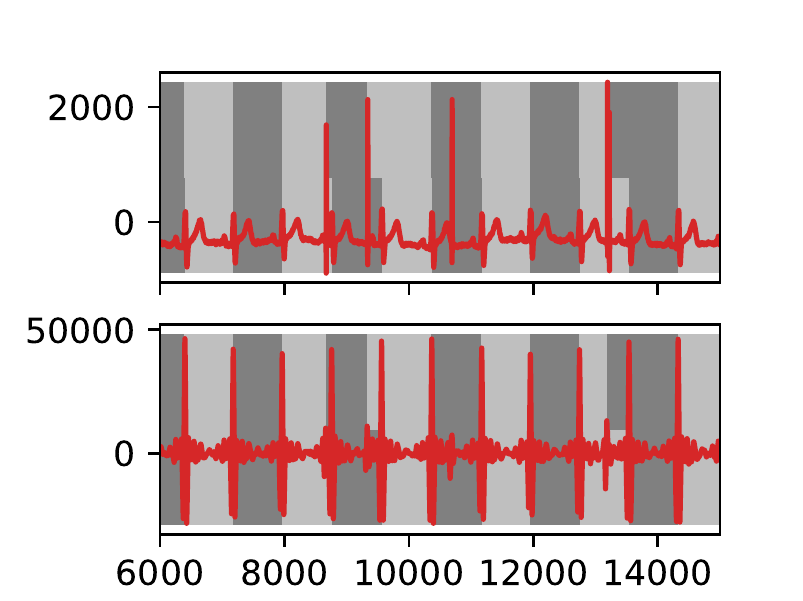}
  \caption{Comparison of periods detected in the Steps~\ref{pds-signalpd} and~\ref{pds-xcorrpd}}
  \label{fig:pds-corr}
\end{figure}

A comparison of the periods detected by the simple peak detector from Step~\ref{pds-signalpd} and the cross-correlating period detector from Step~\ref{pds-xcorrpd} can be seen in Figure~\ref{fig:pds-corr}.
The top graph shows the input to the simple peak detector, the bottom graph shows the cross-correlation; the gray boxes in the top half of the backgrounds represent the segments inferred by the simple peak detector, those in the bottom half represent those found by the cross-correlating period detector.
Notice how glitches in the input signal easily manage to confuse the simple peak detector while the cross-correlating period detector is robust to such perturbations.

\beginnew
\section{Comparison with other methods}\label{sec:comp}

In this section we perform some comparative evaluation of other methods in order to highlight in particular the utility of phase classification via convolutional neural networks for anomaly detection.
We consider two classes of methods: distance-based approaches employing various types of Euclidean distance comparison (cf.~Sections~\ref{sec:comp-self-sim} and~\ref{sec:comp-dist-cls}) and one-step ahead forecasting (cf.~Section~\ref{sec:comp-lstm}).
In the first class of comparison we demonstrate that even in a phase classification framework, simply comparing the Euclidean-type norm of segments of the underlying signals is less suited for capturing the essence of complex and noise corrupted data.
In the second class of comparison we show that even with the highly complex parametrisation of LSTMs, anomaly detection based on one-step ahead prediction is prone to false positive results.
Since the ECG dataset exhibits the highest level of diversity and is thus most difficult to treat among all example datasets introduced in Section~\ref{sec:example datasets}, for this demonstration we only evaluate the reference methods on this dataset.

\subsection{Self-similarity approach}\label{sec:comp-self-sim}
One way of detecting anomalies in periodic signals is to take a sliding window of roughly one period length, normalise it, and look for a similar segment in the data preceding the window by e.g.~one to two periods.
A threshold is then used to determine whether the considered window is similar enough to one of the preceding segments.
This principle is used in the so-called matrix profiles approach, cf.~e.g.~\cite{YZU16}.
No training data is used in this method and thus no particular characteristics of the normal data themselves are employed during the anomaly detection.
The only point where training data can be useful in this approach is to determine the similarity threshold mentioned above, choosing it so as to avoid having too many false positives on non-anomalous data.

\subsubsection{Method}
Formally, if $\{ X_{t} \}_{t}$ is the input signal and $T$ is the window length, normalise each segment $X^{(\tau)}:= \{ X_{\tau + t} \}_{t = 0, \ldots, T - 1}$ for $\tau \ge 0$ in an analogous manner to that of Section~\ref{subsec:normalise} and denote by $\tilde{X}^{(\tau)}$, $\tau \ge 0$, the respective normalised segments.
Now choose a minimum shift $d_{\min}$ and a maximum shift $d_{\max}$ and compute for each $\tau \ge d_{\max}$
\[
     a_{\tau}
  := \min_{\tau' = \tau - d_{\max}, \ldots, \tau - d_{\min}} \lVert \tilde{X}^{(\tau')} - \tilde{X}^{(\tau)} \rVert^{2}
\]
where $\lVert \tilde{X}^{(\tau')} - \tilde{X}^{(\tau)} \rVert$ denotes the Euclidean distance of $\tilde{X}^{(\tau')}$ to $\tilde{X}^{(\tau)}$, i.e.,
\[
    \lVert \tilde{X}^{(\tau')} - \tilde{X}^{(\tau)} \rVert^{2}
  = \sum_{t = 0}^{T - 1} \lVert \tilde{X}^{(\tau')}_{t} - \tilde{X}^{(\tau)}_{t} \rVert^{2}_{\mathbb{R}^d}
  .
\]
$a_{\tau}$ is called the \emph{self-dissimilarity} of $\{ X_{t} \}_{t}$ at time $\tau$.

Now depending on the type of problem, there are two ways to decide whether a signal is anomalous:
If the task is one of Type~A described in Section~\ref{sec:task}, the average self-dissimilarity of the test signal is computed and compared against some threshold which can for instance be determined by the average self-dissimilarities of the training signals.
If on the other hand the task is one of Type~B described Section~\ref{sec:task}, a threshold is chosen close to the maximum self-dissimilarity of the known normal part of the signal and the self-dissimilarity for the remaining part of the signal is compared against this threshold.

\subsubsection{Results}
\begin{figure}[htbp!]
  \centering
  \begin{subfigure}{\linewidth}
    \includegraphics[width=.95\linewidth]{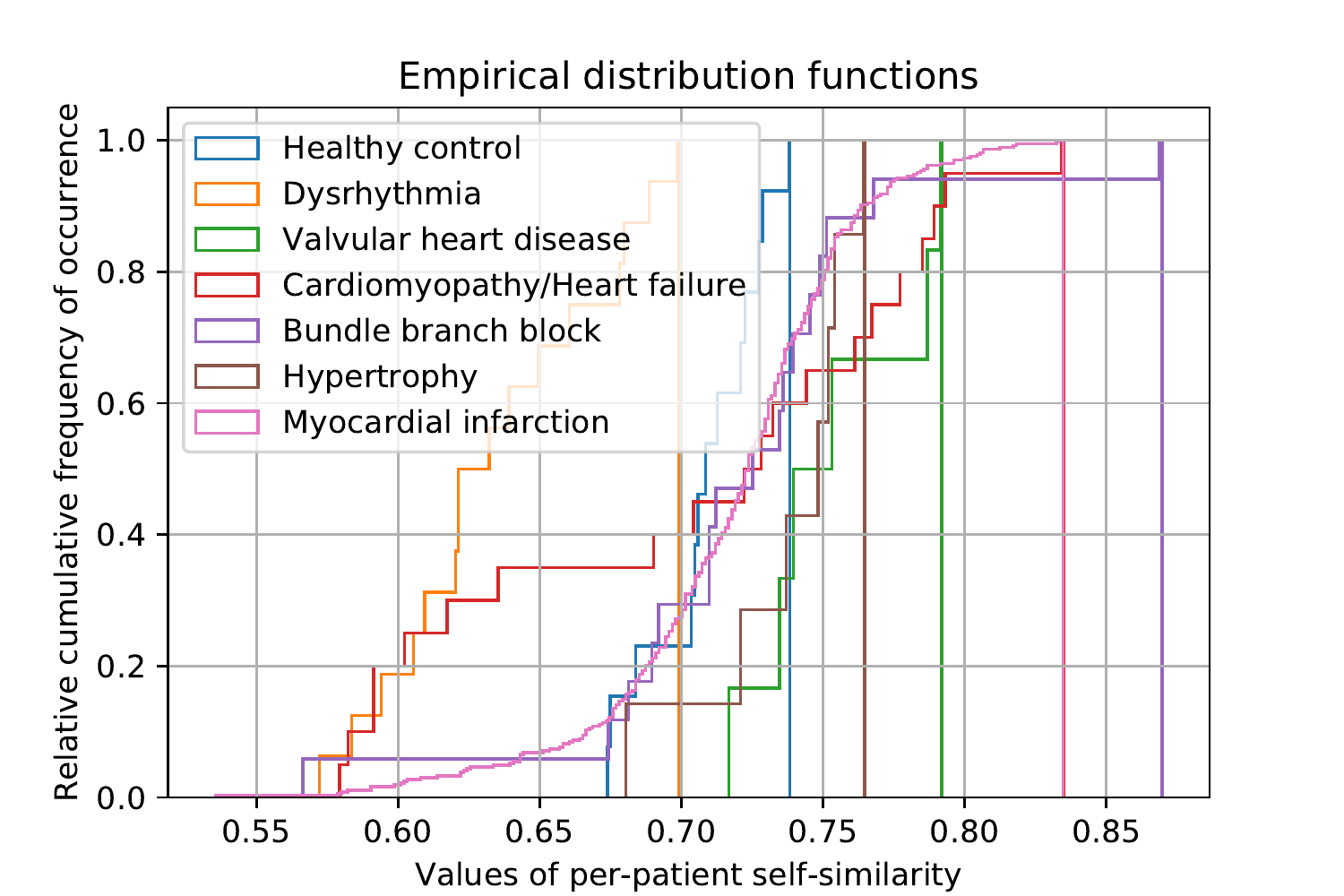}
  \caption{\beginnew Distribution of per-patient self-similarity evaluated on test patients from different categories\endnew}
  \label{fig:hist-self-sim}
  \end{subfigure}
  \begin{subfigure}{\linewidth}
    \includegraphics[width=.95\linewidth]{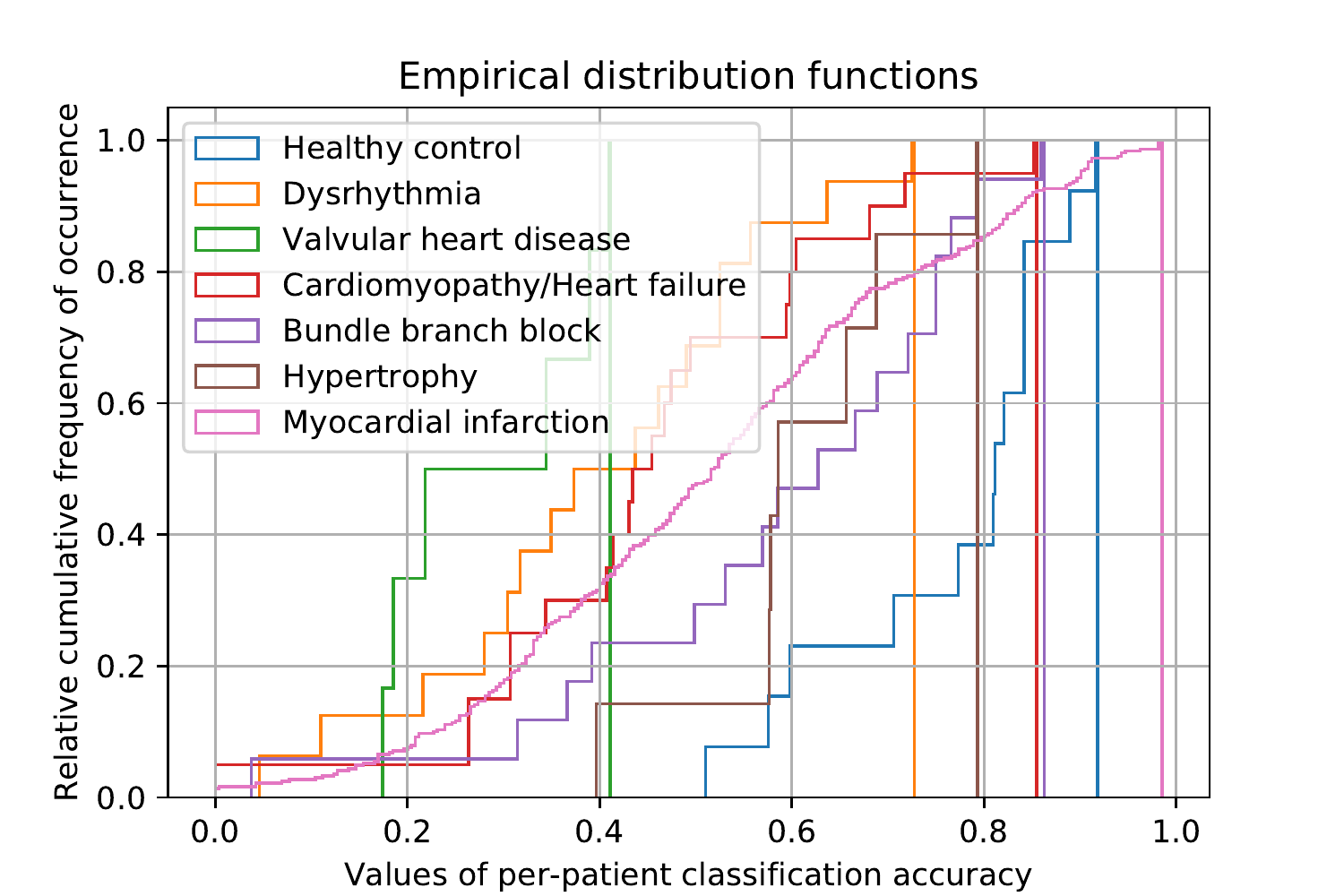}
  \caption{\beginnew Distribution of per-patient distance-based classification accuracy evaluated on test patients from different categories\endnew}
  \label{fig:hist-dist}
  \end{subfigure}
  \begin{subfigure}{\linewidth}
    \includegraphics[width=.95\linewidth]{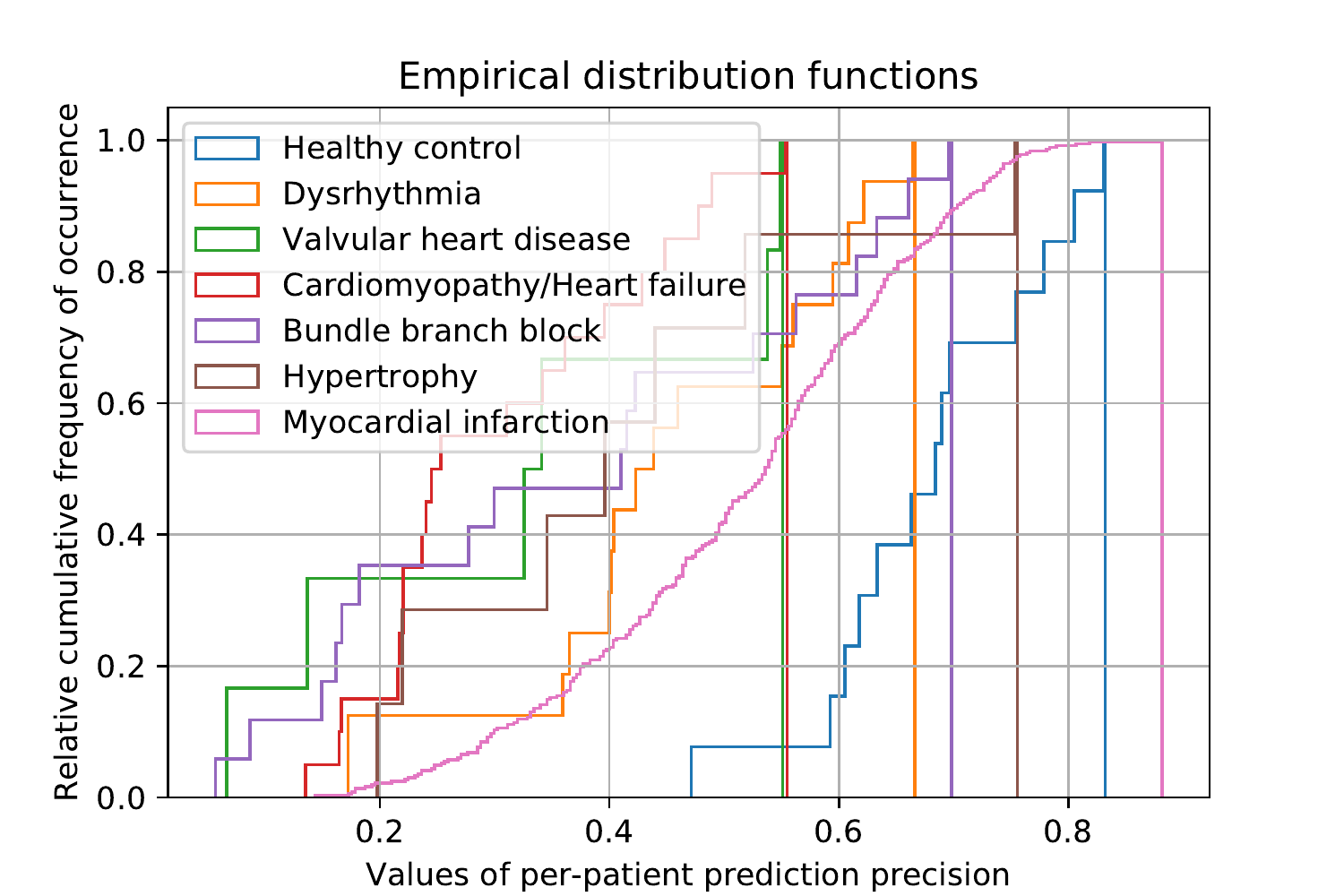}
  \caption{\beginnew Distribution of per-patient forecasting precision evaluated on test patients from different categories\endnew}
  \label{fig:hist-forecast}
  \end{subfigure}
  \caption{\beginnew Distribution of per-patient values for different comparison algorithms\endnew}
  \label{fig:hist-comp}
\end{figure}

For the sake of comparison, we evaluate the performance of the self-similarity-based approach on the ECG database in a similar manner as in our main result in Section~\ref{sec:result-cardio} and first transform the self-dissimilarity computed as described above into a self-similarity rating via the transformation $x \mapsto 1 / (x + 1)$.
We then average the self-similarities for each recording and plot the distributions of these averages grouped by disease.
This plot is shown in Figure~\ref{fig:hist-self-sim}.
One can clearly see that, apart from patients of the category `dysrhythmia' which have the lowest self-similarity, this approach does not manage to produce any separation of ill from healthy patients.

\subsection{Distance-based phase classification}\label{sec:comp-dist-cls}
A distance analysing method similar to that of Section~\ref{sec:comp-self-sim} but more closely related to our main approach is to compute reference windows for the different phases of non-anomalous signals and use these to classify the corresponding segments of the other signals by assigning the class whose reference window has maximum similarity.
Basically, this is the same method as our phase classification scheme but with the classifier neural network replaced by a simple nearest reference classifier.

\subsubsection{Method}
For this method the same data pre-processing with respect to a chosen number of classes $n_{0}$ as described in Section~\ref{sec:pre-processing} is applied to training, validation, and test signals.

For the training dataset $\mathcal{X}$ consisting of normalised segments $\tilde{X}^{\theta}$ labelled as belonging to class $\theta$ for $\theta = 0, \ldots, n_{0} - 1$ (recall the set $\mathcal{X}$ in Figure~\ref{fig:block} for both types A and B), we compute for each class of phase $\theta$ the seasonal averages
\[
     \tilde{X}^{\theta, \operatorname{mean}}
  := \frac{n_{0}}{\# \mathcal{X}} \sum_{\tilde{X}^{\theta} \in \mathcal{X} \text{ labelled } \theta} \tilde{X}^{\theta}.
\]
The classification of a normalised segment $\tilde{X}^{(m)} = \{ \tilde{X}^{(m)}_{t} \}_{t = 0, \ldots, T - 1}$ (labelled $m \bmod n_{0}$) of a test (or validation) signal $\{ X_{t} \}_{t}$ is now performed by computing
\[
  \argmin_{\theta = 0, \ldots, n_{0} - 1} \lVert \tilde{X}^{(m)} - \tilde{X}^{\theta, \operatorname{mean}} \rVert
\]
where again $\lVert \cdot \rVert$ denotes the Euclidean norm as described in Section~\ref{sec:comp-self-sim}.

The remaining part of anomaly detection is performed as in Section~\ref{sec:anom-detect}.

\subsubsection{Results}
To evaluate the performance of the distance-based phase classier on the ECG database, just as in our main result in Section~\ref{sec:result-cardio} we record the classification accuracy on the different types of heart disease and analyse the distribution of the per-patient classification accuracy grouped by the corresponding disease.
The separation into training, validation, and test data is the same as in our main experiment on the ECG database (cf.~Section~\ref{sec:cardio-input}).
For the number of classes, a setting of $n_{0} = 6$ shows the best results, which conforms to our model selection result (cf.~optimal $n_{0}$ presented in Section~\ref{sec:result-cardio}).
The average validation accuracy amounts to $79\%$.
The per-disease average classification accuracy is evaluated in Table~\ref{tab:acc-dist}.
\begin{table}[htbp!]
  \caption{\beginnew Results of per-disease classification accuracy\endnew}
  \label{tab:acc-dist}
  \centering
  \begin{tabular}{rrrr}
       \hline
       \multicolumn{1}{c}{\textbf{Disease}} & \multicolumn{1}{c}{\textbf{Classification}} \\ & \multicolumn{1}{c}{\textbf{Accurracy}}
    \\ \hline
       Valvular heart disease        & $28\%$
    \\ Dysrhythmia                   & $37\%$
    \\ Cardiomyopathy/Heart failure  & $44\%$
    \\ Myocardial infarction         & $51\%$
    \\ Bundle branch block           & $59\%$
    \\ Hypertrophy                   & $60\%$
    \\ \hline
       Healthy control               & $78\%$
    \\ \hline
  \end{tabular}
\end{table}
A plot of the distribution of per-patient classification accuracy evaluated on test patients from different categories is shown in Figure~\ref{fig:hist-dist}.
As can be seen from both the table and the plot when compared to the results of our approach (cf.~Figure~\ref{fig:Figure-hist} and Table~\ref{tab:res-ECG}), the convolutional classifier neural network delivers generally better classification performance with far better results being obtained on the healthy control patients.
In particular, we see that the blue line representing the healthy test patients in Figure~\ref{fig:Figure-hist} is located much closer to the bottom right corner than in Figure~\ref{fig:hist-dist}, indicating better modelling of the normal data by our convolutional classifier neural network.
An anomaly detection using the distance-based classifier thus would have a higher false positive rate than one using the convolutional neural network for classification when achieving comparable detection performance.
For instance, according to Figure~\ref{fig:hist-dist}, if we use the average validation accuracy of $79\%$ as the threshold value as discussed in our main result in Section~\ref{sec:result-cardio}, a similar detection rate in most of the ill categories but a higher false positive rate of $38\%$ on healthy test patients will be achieved compared to the result of our approach ($25\%$ on healthy test patients, cf.~Section~\ref{sec:result-cardio}).

\subsection{Long short-term memory predictor approach}\label{sec:comp-lstm}
As described in Section~\ref{sec:lstm}, one can use a long short-term memory unit (LSTM) to predict the signal one time step ahead, then use a threshold on the difference of this prediction to the actual signal to decide whether the signal behaves as expected or should be considered anomalous.
We choose to demonstrate this method in preference to the statistical forecasting approaches mentioned in Section~\ref{sec:ARIMA}, as no further adjustment to the method is needed for handling problems of Type~A described in Section~\ref{sec:task} and, more importantly, the forecasting performance of LSTMs on data with complex patterns has been shown to be better than that of linear models in general.

\subsubsection{Method}
Since LSTMs are a somewhat complex type of recurrent neural network, we will not describe their construction here and instead refer the reader to the literature on the subject, e.g.~\cite{HS97}.
In our treatment of the ECG database, we use an LSTM with an input layer size of 15, a hidden layer size of 60, and an output layer size of again 15.
We use a mean-squared-error loss function, discarding the first 200 predictions (four seconds) to allow the LSTM to first align with the given signal.
We use the same ADAM algorithm for the training of the LSTM that we also employed for training our convolutional classifier neural networks with a learning rate of $\gamma = 2^{-10}$ and $2^{10}$ training epochs.
The separation into training, validation, and test data is also the same as in our main experiment on the ECG database (cf.~Section~\ref{sec:cardio-input}).

\subsubsection{Results}
\begin{figure}[htbp!]
  \centering
  \begin{subfigure}{\linewidth}
    \includegraphics[width=.95\linewidth]{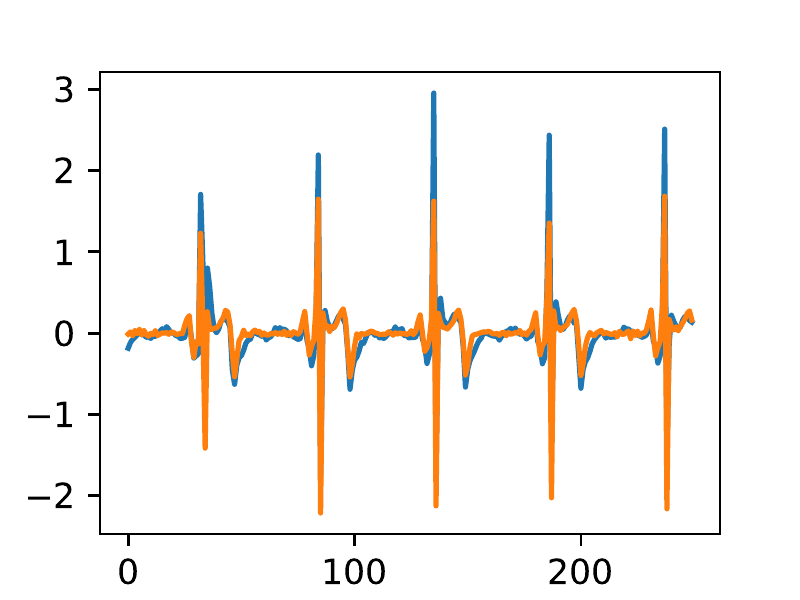}
  \caption{\beginnew Prediction for a healthy patient\endnew}
  \label{fig:pred-healthy}
  \end{subfigure}
  \begin{subfigure}{\linewidth}
    \includegraphics[width=.95\linewidth]{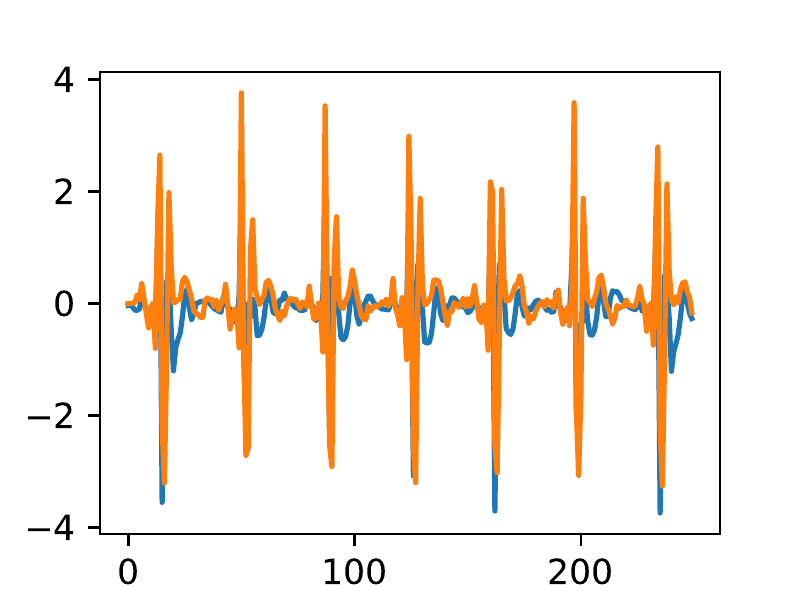}
  \caption{\beginnew Prediction for a patient with myocardial infarction\endnew}
  \label{fig:pred-ill}
  \end{subfigure}
  \caption{\beginnew Example prediction results of LSTM predictor\endnew}
  \label{fig:pred}
\end{figure}

To evaluate the performance of the LSTM on the ECG database, we analyse the distribution of the forecasting precision on the patients coming from the different groups.
A plot of this distribution is shown in Figure~\ref{fig:hist-forecast}.
The measure of performance used here is given by $1 / (MSE + 1)$ where $MSE$ denotes the mean squared error of the predictions on the ECG recording.
This transformation is applied again for the sake of easier comparability with our main result in Section~\ref{sec:result-cardio}.
Using the same measure of performance, the prediction precision evaluated on the training and validation data are $91\%$ and $63\%$ on average, respectively.
The large gap between the training and validation performance suggests the presence of the overfitting phenomenon mentioned in Section~\ref{sec:lstm}, whereas our classifier CNN approach does not suffer from this problem (see the consistency of training and validation accuracy results presented in Section~\ref{sec:result-cardio}).
As presented in Figure~\ref{fig:hist-forecast}, if we choose a threshold value of $63\%$ based on the validation performance as discussed in our main result in Section~\ref{sec:result-cardio}, this will lead to a false positive anomaly detection rate of $31\%$ on healthy test patients, which is higher than that of our approach ($25\%$); at the same time, the detection performance of the LSTM-based detector is lower, with e.g.~only about $75\%$ of the patients labelled `myocardial infarction' (the largest category) being detected as anomalous, compared to the almost $85\%$ of our approach.
Furthermore, for illustration purposes, two examples of the predictions coming from the LSTM are displayed in Figure~\ref{fig:pred}.
%\phantom{\begin{figure}\end{figure}}
%\begin{figure}\end{figure}
Notice that for both patients, the prediction fails to guess the (randomly varying) values at the spikes in the signals correctly.
This (randomly) contributes to the mean squared error and thus results in a weaker separation ability of the anomaly detector.
\endnew

\section*{List of Abbreviations}

\begin{description}
  \item[ARIMA] autoregressive integrated moving average
  \item[ARMA] autoregressive--moving average
  \item[CNN] convolutional neural network
  \item[DC] direct current
  \item[ECG] electrocardiogram
  \item[i.i.d.] independent and identically distributed
  \item[IP] Internet Protocol
  \item[LSTM] long short-term memory
  \item[MSE] mean squared error
  \item[PTB] Physikalisch-Technische Bundesanstalt
  \item[RNN] recurrent neural network
  \item[SARIMA] seasonal autoregressive integrated moving average
  \item[SCADA] supervisory control and data acquisition
  \item[SGD] stochastic gradient descent
\end{description}

\begin{backmatter}

\section*{Availability of data and materials}
  The datasets used for the evaluation of the algorithm are available online at PhysioNet \url{https://physionet.org/physiobank/database/ptbdb/} and the GitHub repository \url{https://github.com/antoine-lemay/Modbus_dataset}.

\section*{Competing interests}
  The authors declare that they have no competing interests.

\section*{Funding}
  The work has been funded by the German Research Center for Artificial Intelligence and the Technical University of Kaiserslautern.

\section*{Authors' contributions}
  LA devolped and implemented the core concepts of the algorithm presented within this manuscript, JA provided refinements and performed data acquisition and generation as well as further supplemental programming, HDS provided further technical knowledge and support.

\section*{Acknowledgements}
  The authors would like to thank the anonymous referees for providing valuable suggestions which helped clarify the exposition of the material.
  Furthermore, the authors would like to thank Dr.-Ing.\ J\"{o}rg Schneider for many fruitful discussions, in particular about the cardiology dataset.

\section*{Authors' information}
LA has a Ph.D.\ in Mathematics, specialises in stochastic processes, stochastic filtering, and machine learning, and is currently working as a senior researcher at the German Research Center for Artificial Intelligence.
JA has a Master's degree in mathematics, specialises in non-commutative harmonic analysis, has a background in digital signal processing and machine learning, and is currently working as a researcher at the German Research Center for Artificial Intelligence.
HDS is the Scientific Director of the Intelligent Networks Research Group at the German Research Center for Artificial Intelligence and head of the Institute for Wireless Communication and Navigation at the Technical University of Kaiserslautern.

%%  After submission of the .TEX file,                     %%
%%  you will be prompted to submit your .BBL file.         %%

% if your bibliography is in bibtex format, use those commands:
\bibliographystyle{bmc-mathphys} % Style BST file (bmc-mathphys, vancouver, spbasic).
\bibliography{bib}      % Bibliography file (usually '*.bib' )

\end{backmatter}
\end{document}